\documentclass[useAMS,usenatbib]{mn2e}

\title[Electron impact excitation of Si II]{Energy levels, radiative rates and electron impact excitation rates for transitions in Si II\thanks{Tables 2 and 5 are available only in the electronic version.}}
\author[K. M. Aggarwal and F. P. Keenan]{Kanti  M.  ~Aggarwal$^{1}$\thanks{E-mail:
 K.Aggarwal@qub.ac.uk(KMA); F.Keenan@qub.ac.uk (FPK)} and Francis  P.   ~Keenan$^{1}$ \\
$^{1}$Astrophysics Research Centre, School of Mathematics and Physics, Queen's University Belfast, Belfast BT7 1NN, Northern Ireland, UK} 
\begin{document}

\date{Accepted 2014 May 1. Received 2014  May 1; in original form 2014 February 28}

\pagerange{\pageref{firstpage}--\pageref{lastpage}} \pubyear{2014}

\maketitle

\label{firstpage}

\begin{abstract}

Energies for the lowest 56 levels, belonging to the 3s$^2$3p, 3s3p$^2$, 3p$^3$, 3s$^2$3d, 3s3p3d, 3s$^2$4$\ell$ and 3s$^2$5$\ell$ configurations of Si II, are calculated using the  {\sc grasp} (General-purpose Relativistic Atomic Structure Package) code. Analogous calculations have also been performed (for up to 175 levels) using the Flexible Atomic Code ({\sc fac}). Furthermore,  radiative rates are calculated  for all E1, E2, M1 and M2 transitions. Extensive comparisons are made with available theoretical and experimental energy levels, and the accuracy of the present results is assessed to be better than 0.1 Ryd. Similarly, the accuracy for radiative rates (and subsequently lifetimes)  is estimated to be better than  20\% for most of the (strong) transitions. Electron impact excitation collision strengths are also calculated, with the Dirac Atomic R-matrix Code ({\sc darc}), over a wide energy range up to 13 Ryd. Finally,  to determine effective collision strengths,  resonances are resolved in a fine energy mesh in the thresholds region. These  collision strengths are  averaged over a Maxwellian velocity distribution and results listed over a wide range of temperatures,  up to 10$^{5.5}$ K. Our data are compared with earlier $R$-matrix calculations and  differences noted, up to a factor of two, for several transitions.  Although scope remains for improvement, the accuracy for our results of collision strengths and effective collision strengths is assessed to be about 20\% for a majority of transitions.

\end{abstract}


\section{Introduction}
Emission lines of Si II have been observed at optical and ultraviolet (UV) wavelengths in a variety of plasmas, such as planetary nebulae, quasars and the interstellar medium -- see for example: \cite*{jch}. Particularly useful for diagnostic purposes are the multiplets 3s$^2$3p $^2$P$^o$ -- 3s3p$^2$ $^4$P and 3s$^2$3p $^2$P$^o$ --  3s3p$^2$ $^2$D at around 2340 and 1810 $\rm \AA$, respectively. Similarly, \cite{hg} have detected two emission lines ($\sim$ 7849 $\rm \AA$) in the magnetic Bp star a Centauri (HD 125823). Recently, \cite{shall} have determined the silicon abundance in the solar atmosphere, based on lines of Si ions, including those of Si II. Many lines of Si II in the 750-6680 ${\rm \AA}$ wavelength range are listed in  the CHIANTI database at {\tt http://www.chiantidatabase.org/}  and the {\em Atomic Line List} (v2.04) of Peter van Hoof at ${\tt {\verb+http://www.pa.uky.edu/~peter/atomic/+}}$. Silicon ions, including Si II, are also important for the studies of fusion plasmas, particularly because amorphous silicon is used for coating the first wall of the devices, such as TEXTOR. \cite*{hub} have measured intensities of several Si II lines in the TEXTOR tokamak in the 290-640 ${\rm \AA}$ wavelength range, belonging to the $n \le$ 5 levels. The importance of data for Si ions has further increased with the developing ITER project.

For  plasma diagnostics and modelling,  atomic data are required for a range of parameters, particularly energy levels, radiative rates (A- values), and excitation rates (or equivalently the effective collision strengths $\Upsilon$). Measured values of energy levels have been compiled by the NIST (National Institute of Standards and Technology) team \citep*{nist} and are available at their  website {\tt http://www.nist.gov/pml/data/asd.cfm}. Theoretical energy levels have been determined by several authors, and the most notable results are those of \cite{sst1} and \cite{mab}. Both these workers have also listed the A- values.
 
Results for collision strengths ($\Omega$) and effective collision strengths ($\Upsilon$) are also available.  \cite{dk1} have reported $\Upsilon$ data for transitions among the lowest 7 levels of the 3s$^2$3p and 3s3p$^2$ configurations  and from the 3s$^2$3p $^2$P$^o_{1/2,3/2}$ ground state levels to higher excited ones, up to 15 (see Table 1). These limited results are based on the $R$-matrix method and are primarily in $LS$ coupling (Russell-Saunders  or spin-orbit coupling), but corresponding data for fine-structure transitions were determined through algebraic transformation. Although calculations for $\Omega$ were performed up to an energy of 10 Ryd, which is fully sufficient to determine $\Upsilon$ up to  T$_e$ = 10$^{4.6}$ K, the range of partial waves adopted by them ($J \le$ 6) was too limited to obtain convergence of $\Omega$, not only for the allowed but also  the forbidden transitions. Furthermore,  \cite{jch}  analysed several observations using their data and  noted  discrepancies with theory by up to a factor of two for several lines. On the other hand, \cite{bald} studied broad emission lines of high luminosity QSOs, including some of  Si II, particularly $\lambda \sim$ 1263 $\rm \AA$ (3s$^2$3p $^2$P$^o$ --  3s$^2$3d $^2$D), 1307 $\rm \AA$ (3s$^2$3p $^2$P$^o$ --  3s3p$^2$ $^2$S) and 1814 $\rm \AA$ (3s$^2$3p $^2$P$^o$ --  3s3p$^2$ $^2$D). For the  1814 $\rm \AA$ line there was a satisfactory agreement between prediction and observation, but the discrepancies for the other two lines were  over an order of magnitude. Therefore, there was a clear need to re-examine the theoretical atomic data.

Subsequently, \cite{sst2} made significant improvements over the atomic data of \cite{dk1}, mainly by extending the range of partial waves up to angular momentum $J$ = 36. Furthermore, he performed calculations in the Breit-Pauli B-spline $R$-matrix (BSRM) approach and reported $\Upsilon$ not only over a wider temperature range (up to log T$_e$ = 5.4 K) but also for {\em all} transitions among 31 levels of the 3s$^2$3p, 3s3p$^2$, 3s3p3d, 3s$^2$4$\ell$, 3s$^2$5s/p/d/f and 3s$^2$6s/p configurations. As a result of these improvements the differences between his values of $\Upsilon$ and those of \cite{dk1} were significant for some  transitions, in both magnitude and behaviour, particularly the allowed ones, such as 1--12 (3s$^2$3p $^2$P$^o_{1/2}$ -- 3s$^2$3d $^2$D$_{3/2}$) and 2 --13 (3s$^2$3p $^2$P$^o_{3/2}$ -- 3s$^2$3d $^2$D$_{5/2}$), as shown in his Fig. 6. However, Tayal included only 2 levels ($^2$D$^o_{3/2,5/2}$) of the 3s3p3d configuration, whereas it generates 23 in total (see Table 1), and some of these lie in between those of the 3s$^2$4$\ell$ and 3s$^2$5$\ell$ configurations, included in his calculations. The omission of these levels affects the resonance structure of $\Omega$ and subsequently calculations of $\Upsilon$. Therefore, there is a scope for improvement as well as  confirmation of  accuracy  of the Tayal results, so that data can be confidently applied to plasma studies.

\cite{mab} performed another calculation adopting the Breit-Pauli $R$-matrix code of \cite*{rm1}. Although a large number of (43) $LS$ terms  were included  in the collisional calculations, they reported values of $\Upsilon$ only at three temperatures {\em and} for 8 transitions from the ground level. More importantly, discrepancies with the corresponding results of \cite{dk1} as well as \cite{sst2} are up to 50\% for several transitions. In general, there is poor agreement among the three $R$-matrix calculations. \cite{mab} speculated that differences with the calculations  of \cite{sst2} could be because of the non-orthogonal orbitals adopted by him. However, \cite{dk1}  adopted orthogonal orbitals, as did \cite{mab}, but differences between the two sets of $\Upsilon$ are still up to 50\% at all temperatures between 5000 and 20,000 K. A closer examination of the collision strengths shown by \cite{sst2} and \cite{mab} for three transitions, namely 1--2 (3s$^2$3p $^2$P$^o_{1/2}$ -- 3s$^2$3p $^2$P$^o_{3/2}$), 2--4 (3s$^2$3p $^2$P$^o_{3/2}$ -- 3s3p$^2$ $^4$P$_{1/2}$) and 2--7 (3s$^2$3p $^2$P$^o_{3/2}$ -- 3s3p$^2$ $^2$D$_{5/2}$), reveals that background values of $\Omega_B$ are {\em lower} in the latter's calculations, particularly at energies below 0.8 Ryd (equivalent to $\sim$ 1.25$\times$10$^5$ K). These differences clearly lead to the lowest values of $\Upsilon$ reported by \cite{mab} for all transitions and at all temperatures. 

\cite{mab} included a larger range of partial waves ($L \le$ 16) in comparison to those of \cite{dk1}, who included only $L \le$ 8, and yet their values of $\Upsilon$ are the {\em lowest}, not only for the allowed (such as 1-- 3, 4 and 6) but also the forbidden (1--2 and 1--5) transitions. Therefore, neither the range of partial waves nor the use of (non) orthogonal orbitals appears to be a convincing cause of the discrepancies in $\Upsilon$. For this reason we have performed another calculation for the important Si II ion and report a complete set of results for energy levels, A- values and $\Upsilon$ for all transitions among 56 levels of the 3s$^2$3p, 3s3p$^2$, 3p$^3$, 3s$^2$3d, 3s3p3d, 3s$^2$4$\ell$ and 3s$^2$5$\ell$ configurations. 

\section{Energy levels}

To  calculate energy levels and A- values we have employed the  multi-configuration Dirac-Fock (MCDF) code,  developed by  \cite{grasp0}. It is a fully relativistic code,  based on the $jj$ coupling scheme, and includes higher-order relativistic corrections arising from the Breit (magnetic) interaction and quantum electrodynamics (QED) effects (vacuum polarisation and Lamb shift).  This code has undergone several revisions by the names GRASP (General-purpose Relativistic Atomic Structure  Package), GRASP92 and GRASP2K, revised by \cite{grasp},  \cite{grasp92} and  \cite{grasp2k,grasp2kk}, respectively. However, the version adopted here has been revised by one of its authors (Dr. P. H. Norrington), is known as GRASP0 and  is freely available at the website {\tt http://web.am.qub.ac.uk/DARC/}. This  provides comparable results with other revised versions and has been extensively applied by ourselves and other workers to a wide range of ions.  Furthermore, as in our earlier work, the option of {\em extended average level} (EAL) has been adopted in which  a weighted (proportional to 2$j$+1) trace of the Hamiltonian matrix is minimised. We note that results obtained for energy levels and A- values are comparable to other options, such as {\em average level} (AL), as demonstrated  by  \cite*{kr35}  for Kr  ions. 

For the determination of atomic structure we have included 14  configurations, namely  (1s$^2$2s$^2$2p$^6$) 3s$^2$3p, 3s3p$^2$, 3p$^3$, 3s$^2$3d, 3s3p3d, 3s$^2$4$\ell$ and 3s$^2$5$\ell$, which generate 56 levels listed in Table 1. Our calculated level energies, obtained  with the inclusion of Breit and QED effects, are given in Table 1 along with the NIST compilations of experimental energies  and  those of \cite{sst2} who adopted the multi-configuration Hartree-Fock (MCHF) code of \cite{mchf}.  Since Si II is only moderately heavy, the contributions of Breit and QED effects are negligible. However, differences with the NIST listings are up to $\sim$ 0.1 Ryd  for some of the levels, particularly 3s3p$^2$ $^2$P$_{1/2,3/2}$ (14 and 15). Additionally, the orderings are also slightly different in a few instances -- see for example levels 17--22. We note here that for some species, such as Al-like Ti X  \citep*{tix}, it is not possible to unambiguously identify the levels, because of strong mixing. However, that is not the case for Si II. The energies obtained by \cite{sst2} are in closer agreement with those of NIST for all levels, because he has used non-orthogonal orbitals. However, his energy level ordering differs with that of NIST in a few instances, such as for 6--7 and 21--22.  Furthermore, using the similar non-orthogonal orbitals, in his earlier calculations \citep{sst1} the energy obtained for the 3s$^2$3p $^2$P$^o_{3/2}$ level is lower by 13\%.  Nevertheless, a clear advantage of this approach is that orbitals can be independently optimised on individual states/levels and thus a higher accuracy can be achieved by minimising the differences with the measured results. However, this approach can only be successfully applied if experimental energies are already available, which is the case for Si II, but not for all  levels, because several are missing from the NIST compilations as seen in Table 1. Since \cite{sst2} performed collisional calculations among experimentally-determined  levels only, he could apply the approach of non-orthogonal orbitals. Nevertheless, most of the scattering codes, including the one adopted here and discussed in section 5, do not have a provision of this option.

\vspace*{0.2 cm}
{\LARGE Table 1 }
\vspace*{0.2 cm}

A most commonly and widely used methodology for an accurate determination of energy levels (and hence subsequent other parameters) is the inclusion of {\em configuration interaction} (CI). This approach has been extensively applied to a wide range of ions by many workers, including ourselves for other Si ions \citep{sions}. Therefore, to assess the effect of additional CI we have performed further calculations with the {\em Flexible Atomic Code} ({\sc fac})  of \cite{fac}. This code, apart from being highly efficient, provides results for energy levels and A- values of comparable accuracy with other atomic structure codes, such as CIV3 (configuration interaction version 3) of \cite{civ3} and GRASP, as shown by \cite{fe15} for three Mg-like ions.

We have  performed three calculations  with the {\sc fac} code with increasing amount of CI,  namely (i) FAC1, which includes the same configurations/levels as in GRASP; (ii) FAC2;  which includes all possible combinations of the 3$\ell$ orbitals (3*3) plus 3s$^2$4$\ell$ and 3s$^2$5$\ell$, generating 164 levels in total;   and finally (iii) FAC3, which  includes a further 11 levels of  3s$^2$6$\ell$, because some of these intermix with those of FAC2.  Energies obtained from these three calculations are listed in Table 1 for comparison with other results.

Our FAC1 energies  agree closely with those from GRASP (within 0.03 Ryd) and the ordering is  also similar, although a few differ slightly, such as  levels 17--22. We also note that small discrepancies in the {\sc grasp} and {\sc fac} energies mainly arise due to the different ways  the  calculations of central potential for radial orbitals and recoupling schemes of angular parts are performed -- see the detailed discussion in the {\sc fac} manual ({\tt {\verb+http://sprg.ssl.berkeley.edu/~mfgu/fac/+}}). Such small differences have also been noted for several other ions.  Inclusion of larger CI in FAC2 lowers the energies, by up to 0.07 Ryd, for some of the levels, such as 12--15, but differences with the NIST compilations remain of up to $\sim$ 0.1 Ryd. Finally, energies obtained in FAC3 are comparable (within 0.01 Ryd) to those from FAC2 indicating that the inclusion of the 3s$^2$6$\ell$ configuration has little effect on the energy levels of Table 1. Furthermore, the energy for  level 2 (3s$^2$3p $^2$P$^o_{3/2}$) has become worse than in FAC1 or GRASP, and hence there in no overall advantage of including more CI than that  already considered in GRASP. We discuss this further below.

\cite{mab} have performed a series of calculations with the atomic structure (AS) code of  \cite{as}. Apart from adopting different optimisation procedures (including the minimisation with observed energies) they have included {\em extensive} CI with up to $n$ = 5 orbitals in 44 configurations. They have also opened the $n$ = 2 shell (frozen in our work with GRASP and FAC)  to account for the core-valence correlation. However, none of their  nine calculations yielded accurate energies for the lowest 15 levels. Differences with the measurements for all sets of energies are over 15\% for some levels or others, as shown in their Table 2. It is clear that inclusion of a large amount of CI is not helpful for the accurate determination of Si II energy levels.

Apart from the inclusion of CI, another possibility of improving the accuracy of wave functions is to add correlation effects through pseudo orbitals, i.e. all orbitals need not be spectroscopic as has been the case in the  calculations described above with the GRASP, FAC  and AS codes. This is a normal practice in standard $R$-matrix calculations for $\Omega$, for which input wave functions are generated through the CIV3 code, but all orbitals are orthogonal.  Therefore, \cite{pld} adopted this approach but their energy levels still differ by up to 6\% with the measurements. Additionally, a disadvantage of this approach is the presence of unphysical pseudo resonances,  because the corresponding eigenstates are not  included in the calculations of $\Omega$ -- see for example, Fig. 1 of \cite{ah1}. If these pseudo resonances  are not properly removed then the subsequent results for $\Upsilon$ may be significantly overestimated -- for examples see Figs. 1 -- 6 of \cite{fexv}. Therefore, scope remains for improvement over our energy levels listed in Table 1, but keeping in mind our further calculations for more important parameters ($\Omega$ and $\Upsilon$), the accuracy achieved should be satisfactory for a majority of the levels. 

\section{Radiative rates}

Using the GRASP code we have calculated  A- values for four types of transitions, namely electric dipole (E1), electric quadrupole (E2), magnetic dipole (M1) and  magnetic quadrupole (M2). In general, E1 transitions  dominate in magnitude, but sometimes other types of transitions are also important and hence are (preferably)  required for  a  complete plasma model. The absorption oscillator strength ($f_{ij}$) and radiative rate A$_{ji}$ (in s$^{-1}$) for any type of  transition $i \to j$ are related by the following expression:

\begin{equation}
f_{ij} = \frac{mc}{8{\pi}^2{e^2}}{\lambda^2_{ji}} \frac{{\omega}_j}{{\omega}_i} A_{ji}
 = 1.49 \times 10^{-16} \lambda^2_{ji}    \frac{{ \omega}_j}{{\omega}_i} A_{ji}                                 
\end{equation}
where $m$ and $e$ are the electron mass and charge, respectively, $c$ velocity of light,  $\lambda_{ji}$  the transition energy/wavelength in $\rm \AA$, and $\omega_i$ and $\omega_j$  the statistical weights of the lower ($i$) and upper ($j$) levels, respectively. However, the relationships between oscillator strength f$_{ij}$ (dimensionless) and the line strength S (in atomic unit, 1 a.u. = 6.460$\times$10$^{-36}$ cm$^2$ esu$^2$) with the A- values are different for different types of transitions -- see Eqs. (2--5) of  \cite{tixix}.

\vspace*{0.2 cm}
{\LARGE Table 2 }
\vspace*{0.2 cm} 

In Table 2 we list transition (energies) wavelengths ($\lambda$, in $\rm \AA$), radiative rates (A$_{ji}$, in s$^{-1}$), oscillator strengths (f$_{ij}$, dimensionless), and line strengths (S, in a.u.) for all 460 electric dipole (E1) transitions among the 56 levels of  Si II.  The A-,  f-  and S- values have been calculated in both Babushkin and Coulomb gauges, i.e.  the length and velocity forms in the widely used non-relativistic nomenclature. However,  in Table 2  results are listed in the length form alone, because  the velocity form is generally  considered to be comparatively less accurate.  Nevertheless,  we will discuss  later the velocity/length  form ratio, as this provides some assessment of  the accuracy of the results. Also note that the {\em indices} used  to represent the lower and upper levels of a transition correspond to those in Table 1. Apart from the above E1 transitions, there  are 733 electric quadrupole (E2), 567  magnetic dipole (M1), and 574 magnetic quadrupole (M2) transitions  among the same 56 levels. However, for these only the A-values are listed in Table 2, because these are the ones required for plasma modelling. The corresponding results for f-  values can be easily obtained through Eq. (1). Similarly, if required, corresponding S- values can be obtained through Eqs. (2--5) of  \cite{tixix}.

In Table 3 we compare our f- values from GRASP with those of \cite{sst1} and \cite{mab} for transitions among the lowest 30 levels of Table 1.  \cite{mab} have listed several sets of f- values, but those included in Table 3 correspond to their `recommended' results, which are based on the averages of a variety of theoretical and experimental (to be discussed later) values. These authors reported f- values  only for transitions among the lowest 15 levels, but  many are missing from the work of \cite{sst1}. Most of the missing transitions are {\em weak} (i.e. f $\sim$ 10$^{-5}$ or even less), but some are rather strong, such as 6--22 (f = 0.125), 7--21 (f = 0.120) and 12--22 (f = 0.734). This is because he included the 3s$^2$4f (21--22) levels in his  calculations for collisional data \citep{sst2}, but not for the radiative rates. Nevertheless, the absence of f- (or A-) values for these transitions may affect the modelling of plasmas. 

\vspace*{0.2 cm}
{\LARGE Table 3 }
\vspace*{0.2 cm}

Among the {\em common} transitions, f- values differ by up to a factor of two for a few, such as 2 -- 20, 8 -- 10/11 and 2 -- 24. These transitions are comparatively strong (0.15 $\le$ f $\le$ 0.89) and therefore better agreement is expected. However, such discrepancies  do arise with varying amounts of CI (and methods/codes) as discussed in detail by \cite{mab} -- see their Figs. 2 and 3, and particularly Table 3. Similarly, discrepancies for some weak transitions (such as 1-- 4, 2 -- 4/7 and 4 -- 11) are up to an order of magnitude. Weaker transitions are more susceptible to varying amount of CI, due to cancellation or additive effects, and hence such discrepancies are very common -- see also \cite{oh}. Overall, there is a good agreement among all calculations for most transitions.

For a few transitions, measured  f- values are also available  -- see \cite{mab} and references therein, but there is significant scatter among these. Therefore, the accuracy of the radiative data cannot be assessed by comparison with experimental results. However, another criterion normally used to assess the accuracy of  f- or A- values is to compare the ratio (R) of their  velocity and length forms. This should ideally be close to unity but often is not \citep*{fe15},  because the two formulations are not exactly the same.  Similarly, different calculations with differing amount of CI may yield R closer to unity but strikingly  different f- values in magnitude, as already stated.  Nevertheless, we  include in Table 3  the ratio of the velocity and length forms obtained in our calculations with GRASP.

For most (comparatively strong) transitions listed in Table 3 the ratio R is  within $\sim$ 20\% of 1.0. However, there are exceptions. For example, for the 1-- 8 (3s$^2$3p $^2$P$^o_{1/2}$ -- 3s$^2$4s $^2$S$_{1/2}$) and 2 -- 8 (3s$^2$3p $^2$P$^o_{3/2}$ -- 3s$^2$4s $^2$S$_{1/2}$) transitions, there is no discrepancy among the f- values from GRASP, MCHF and AS, but R is 0.56. On the other hand, for the 8 -- 10/11 (3s$^2$4s $^2$S$_{1/2}$ -- 3s$^2$4p $^2$P$^o_{1/2,3/2}$) transitions, R is 0.92 but the f- values differ by almost a factor of two. Therefore, as explained earlier, this assessment cannot be rigorously applied, but the overall accuracy of the listed results for strong transitions appears to be satisfactory. For weak(er) transitions, R is up to 10,000, the discrepancies among f- values are larger, and hence the accuracy is lower. For similar reasons, the accuracy for the  E2, M1 and M2 data is also lower, because most of these are weaker in comparison to the E1 transitions.  Finally, as for energy levels, we have calculated A- and f- values from FAC also, and for most (strong) transitions the agreement with our GRASP results is within $\sim$ 20\%.

\section{Lifetimes}

The lifetime $\tau$ for a level $j$ is determined  as follows:

\begin{equation}  {\tau}_j = \frac{1}{{\sum_{i}^{}} A_{ji}}.  
\end{equation} 

Its measurement  can directly lead to the  assessment of  accuracy of the A- values, particularly if a single transition dominates. Therefore, in Table 1 we have also listed our calculated lifetimes. As already stated, A- values for E1 transitions  generally dominate, but for completeness we have also  included the contributions from E2, M1 and M2.  Their  inclusion is particularly important for those levels which do not connect via E1 transitions. \cite*{cur} have measured $\tau$ corresponding to the 3s$^2$3p $^2$P$^o_{3/2}$ -- 3s$^2$4s $^2$S$_{1/2}$ transition to be 0.91$\pm$0.04 ns, which compares well with our result of 0.996 ns. Correspondingly, the measured f- value for this 2 -- 8 transition is 0.130$\pm$0.006 and our calculated result is 0.150. However, determination of this f- value by several authors differ by up to  a factor of two, as shown in Table 1 of \cite{cur}. Similarly, their measured $\tau$ for the 3s$^2$5s $^2$S$_{1/2}$ level is 1.99$\pm$0.12 ns, in close agreement with our result of 1.865 ns and the calculation of 2.501 ns by  \cite*{hos}. However, in a private correspondence to \cite{cur}, Hibbert estimated $\tau$ to be 2.36 ns from improved calculations, and this improves the agreement  between the theoretical results. Finally, we also note that for this level apart from the dominant 1--16 and 2--16 (3s$^2$3p $^2$P$^o_{1/2,3/2}$ -- 3s$^2$5s $^2$S$_{1/2}$) E1 transitions, the contributions of the 10 -- 16 and 11 -- 16  (3s$^2$4p $^2$P$^o_{1/2,3/2}$ -- 3s$^2$5s $^2$S$_{1/2}$) E1 are also significant, as mentioned by \cite{cur}.

The other measurements of $\tau$ available are by \cite*{cal} for the 3s3p$^2$ $^4$P$_{1/2,3/2,5/2}$ levels, which are 104$\pm$16, 811$\pm$77 and 406$\pm$33 $\mu$s, respectively. However, our corresponding theoretical results for these levels (251, 8004 and 3391 $\mu$s) are overestimated by up to an order of magnitude. The dominant contributing E1 transitions for these levels are 2 -- 3, 2 -- 4 and 2 -- 5 for which the f- values are 1.15$\times$10$^{-6}$, 1.17$\times$10$^{-7}$ and 4.41$\times$10$^{-7}$, respectively, i.e. all transitions are very {\em weak} and for these there are large variations among different calculations, as discussed in section 3. The other partial reason for the large discrepancies is the comparative inaccuracy of our calculated energy for level 2 (3s$^2$3p $^2$P$^o_{3/2}$), as discussed in section 2 and shown in Table 1. Finally, \cite{bash} have made beam foil measurements of $\tau$ for the (3s$^2$) 4p and 4f levels. Their measured $\tau$ of 9.1$\pm$0.5 and 3.4$\pm$0.3 ns compare well with our corresponding calculations of 8.9 and 3.9 ns.

  \section{Collision strengths}

For the  calculations of $\Omega$, we have employed the {\em Dirac Atomic $R$-matrix Code} ({\sc darc}) of P. H. Norrington and I. P. Grant, available at the website {\tt http://web.am.qub.ac.uk/DARC/}. This  is a relativistic version of the standard $R$-matrix code. Since the code  is based on the $jj$ coupling scheme (i.e. including fine-stucture) the size of the Hamiltonian increases in a calculation, but it (generally) leads  to higher accuracy  (for $\Omega$ and subsequently $\Upsilon$), especially for transitions among the {\em fine-structure} levels of a state, because resonances through the energies of degenerating levels are also taken into  account.  However, because Si II is only moderately heavy and degeneracy among its levels is not large (see Table 1) the results obtained should be comparable with those from the standard $R$-matrix code \citep{rm1}, provided the input parameters are similar.

The $R$-matrix radius adopted for  Si II is 36.8 au, and 56  continuum orbitals have been included for each channel angular momentum in the expansion of the wave function. This large expansion is computationally more demanding as the corresponding size of the Hamiltonian matrix is 15,154. However, it  allows us to compute $\Omega$ up to an energy of  13 Ryd,  or equivalently  values of effective collision strength $\Upsilon$ (see section 6)  up to T$_e$ = 6.0 $\times$10$^{5}$ K, more than an order of magnitude higher than  the temperature of maximum abundance in ionisation equilibrium, i.e. 20,000 K  \citep*{pb}.  The maximum number of channels for a partial wave is 268 and  all partial waves with angular momentum $J \le$ 40 are included.  

Inclusion of a large range of partial waves ensures convergence of $\Omega$ for all forbidden and inter-combination transitions, and at all energies. However, for some allowed transitions  a larger range is preferable because $\Omega$ are not fully converged, particularly at higher energies. We demonstrate the variation of $\Omega$ in Fig. 1 (a, b and c) at three energies (2, 6 and 10 Ryd) and for three transitions, namely 1 -- 12 (3s$^2$3p $^2$P$^o_{1/2}$ -- 3s$^2$3d $^2$D$_{3/2}$), 2 -- 13 (3s$^2$3p $^2$P$^o_{3/2}$  -- 3s$^2$3d $^2$D$_{5/2}$) and 2 -- 15 (3s$^2$3p $^2$P$^o_{3/2}$ -- 3s3p$^2$ $^2$P$_{3/2}$). It may be seen in these figures that $\Omega$ have (almost) converged up to an energy of 6 Ryd, but not at the higher values. To improve the accuracy of $\Omega$ at such energies, i.e. to account for higher neglected partial waves, we have included a top-up, based on the Coulomb-Bethe  approximation of    \cite{ab}. Furthermore, we have also included such contributions for  forbidden transitions,  based on  geometric series, but these are  small. 

\vspace*{0.2 cm}
{\LARGE Figures 1a,b,c }
\vspace*{0.2 cm}

In Table 4 we list our values of $\Omega$ for  transitions from  the levels of the ground state (3s$^2$3p $^2$P$^o_{1/2,3/2}$) of  Si II at three energies of 2, 6 and 10 Ryd. The  indices used  to represent the levels of a transition correspond to those defined in Table 1. Similar results are not available from the work of \cite{mab}, but \cite{sst2} has reported values of $\Omega$  in the 2--10 Ryd energy range. In Fig. 2 we compare our results of $\Omega$ with those of \cite{sst2} for three {\em forbidden} transitions, namely 1 -- 10 (3s$^2$3p $^2$P$^o_{1/2}$ -- 3s$^2$4p $^2$P$^o_{1/2}$), 1 -- 21 (3s$^2$3p $^2$P$^o_{1/2}$ -- 3s$^2$4f $^2$F$^o_{7/2}$) and 2 -- 11 (3s$^2$3p $^2$P$^o_{3/2}$ -- 3s$^2$4p $^2$P$^o_{3/2}$). For all these (and many other)  transitions, $\Omega$ have fully converged within our partial waves range of $J \le$ 40. \cite{sst2} included partial waves with $J \le$ 36, comparable to ours and hence fully sufficient, yet his values of $\Omega$ are consistently lower by up to $\sim$ 23\% for all transitions. On the other hand, his $\Omega$ for {\em allowed} transitions are consistently higher by up to $\sim$ 25\%, as shown in Fig. 3 for three, namely 1 -- 12 (3s$^2$3p $^2$P$^o_{1/2}$ -- 3s$^2$3d $^2$D$_{3/2}$), 2 --13 (3s$^2$3p $^2$P$^o_{3/2}$ -- 3s$^2$3d $^2$D$_{5/2}$) and 2 -- 15 (3s$^2$3p $^2$P$^o_{3/2}$ -- 3s3p$^2$ $^2$P$_{3/2}$). For all these transitions, $\Omega$ have fully converged up to an energy of 6 Ryd as shown in Fig. 1, and yet the Tayal results for $\Omega$ are consistently higher at all energies. For these transitions the f- values from our GRASP and his MCHF \citep{sst1} calculations agree to better than 10\%, as shown in Table 3. However, in the scattering calculations \cite{sst2} has adopted slightly different wavefunctions, and therefore the f- values obtained may not be the same as listed in Table 3. In spite of these being comparatively strong transitions, the f- values do show significant variations as shown in Table 3 of \cite{mab}. Therefore, if the f- values in the later calculations \citep{sst2} are   higher than so will be the values of $\Omega$, but it cannot be confirmed with the limited information available.

\vspace*{0.5 cm}
{\LARGE Table 4 }

\vspace*{0.2 cm}
{\LARGE Figures 2 and 3 }
\vspace*{0.2 cm}

\section{Effective collision strengths}

Apart from  energy levels and radiative rates, excitation and de-excitation rates are required for plasma modelling, which are determined from the collision strengths ($\Omega$). However, as already shown by \cite{dk2}, \cite{sst2} and \cite{mab}, $\Omega$ does not vary smoothly within  the thresholds region, because of the closed-channel (Feshbach) resonances, especially for (semi) forbidden transitions. Such resonances need to be resolved  in a fine energy mesh  to accurately account for their contribution. In most astrophysical (and fusion) plasmas, electrons have a  {\em Maxwellian} distribution of velocities, and thereofore an averaged value, known as {\em effective} collision strength ($\Upsilon$) is required, i.e. 

\begin{equation}
\Upsilon(T_e) = \int_{0}^{\infty} {\Omega}(E) \, {\rm exp}(-E_j/kT_e) \,d(E_j/{kT_e}),
\end{equation}
where $k$ is the Boltzmann constant, T$_e$  electron temperature in K, and E$_j$  the electron energy with respect to the final (excited) state. Once the value of $\Upsilon$ is
known the corresponding results for the excitation q(i,j) and de-excitation q(j,i) rates can be easily obtained from the following equations:

\begin{equation}
q(i,j) = \frac{8.63 \times 10^{-6}}{{\omega_i}{T_e^{1/2}}} \Upsilon \, {\rm exp}(-E_{ij}/{kT_e}) \hspace*{1.0 cm}{\rm cm^3s^{-1}}
\end{equation}
and
\begin{equation}
q(j,i) = \frac{8.63 \times 10^{-6}}{{\omega_j}{T_e^{1/2}}} \Upsilon \hspace*{1.0 cm}{\rm cm^3 s^{-1}},
\end{equation}
where $\omega_i$ and $\omega_j$ are the statistical weights of the initial ($i$) and final ($j$) states, respectively, and E$_{ij}$ is the transition energy. Depending on the type of transition {\em and} temperature, the contribution of resonances may greatly enhance the values of $\Upsilon$ over those of the background  collision strengths ($\Omega_B$).  Since $\Upsilon$ for Si II are most important  for up to T$_e$ = 20,000 K ($\sim$ 0.13 Ryd), the contribution of resonances is very significant  for most  transitions. In addition,  values of $\Omega$ should  be calculated over a wide energy range (above thresholds)  to obtain convergence of the integral in Eq. (3), as demonstrated in Fig. 7 of  \cite{ni11}. For this reason we have calculated values of $\Omega$ up to an energy of 13 Ryd, as discussed in section 5. 

To resolve resonances, we have performed our calculations of $\Omega$ in a narrow energy mesh  of 0.001 Ryd (at over $\sim$ 1500 energies) in the thresholds region. In Figs. 4 -- 6 we show  resonances for three transitions, namely   1 -- 2 (3s$^2$3p $^2$P$^o_{1/2}$ -- 3s$^2$3p $^2$P$^o_{3/2}$), 2 -- 4 (3s$^2$3p $^2$P$^o_{3/2}$ -- 3s3p$^2$ $^4$P$_{3/2}$) and 2 -- 7 (3s$^2$3p $^2$P$^o_{3/2}$ -- 3s3p$^2$ $^2$D$_{5/2}$), which are forbidden, inter-combination and allowed, respectively. Furthermore, these transitions have specifically  been selected because both \cite{sst2} and \cite{mab} have shown resonances for these, and hence will facilitate us in understanding the differences in the corresponding results of $\Upsilon$. For the three transitions, the resonance structures are similar in all calculations, but their magnitude and background values ($\Omega_B$) differ slightly.  While the $\Omega_B$ of \cite{sst2} are on the higher side, those of \cite{mab} are {\em lower}, particularly at energies below 0.8 Ryd.  We would like to stress here that their lower values of $\Omega$ are not due to the lower range of included partial waves ($L \le$ 16), because at these energies $J \le$ 15 are sufficient for convergence, as demonstrated in Fig. 1 {\em and} also confirmed with our similar plots for all three transitions with $J \le$ 10 and $J \le$ 20.

\vspace*{0.2 cm}
{\LARGE Figures 4, 5 and 6 }
\vspace*{0.2 cm}

Our calculated values of $\Upsilon$ are listed in Table 5 over a wide temperature range of log T$_e$ = 3.7 -- 5.5 K, suitable for applications to a wide range  of laboratory and astrophysical plasmas. Since $\Upsilon$ is a slowly varying function of T$_e$, corresponding data at any other temperature within this range can be easily interpolated,  or may be requested from  the first author. In Table 6, we compare our results of $\Upsilon$ with those of \cite{mab}, \cite{sst2} and \cite{dk1} for transitions from the levels of the  ground state 3s$^2$3p $^2$P$^o_{1/2,3/2}$ to higher excited levels (but only up to 3s3p$^2$ $^2$P$_{3/2}$), and at three temperatures of 5000, 10,000 and 20,000 K. These transitions  are the same as reported by \cite{dk1}. For a majority of  transitions, the $\Upsilon$ of \cite{sst2} are the {\em highest}, although they are also lower for a few, such as 1 -- 8/10/11,  and 2 -- 8/11.  The differences between their $\Upsilon$ and the other calculations are up to 50\%. For all these transitions,  the higher values of $\Upsilon$ by Tayal are because of his corresponding higher values of $\Omega$. We would like to stress here that both ourselves and Tayal have included a comparable large range of partial waves and hence $\Omega$ for all these transitions have fully {\em converged}.  

\vspace*{0.2 cm}
{\LARGE Table 5 and Table 6}
\vspace*{0.2 cm}

Since \cite{sst2} has reported $\Upsilon$ for a wider range of transitions and temperatures, we make some more comparisons with our results. Differences between the two sets of data are up to a factor of two (and larger for only a few) for many transitions, such as  3 -- 8/11/12/13/14/15 and 4 -- 8/12/13/14/15. For most cases his $\Upsilon$ values are higher, but are lower for a few. In Fig. 7 we compare our $\Upsilon$ with those of Tayal for three transitions, namely 3s$^2$3p $^2$P$^o_{3/2}$ -- 3s3p$^2$ $^4$P$_{1/2,3/2,5/2}$, i.e. 2 -- 3/4/5, which are very important for diagnostics as stated in section 1. Towards the lower end of the temperature range his $\Upsilon$ are significantly lower (by a factor of 70) for the 2 -- 3 transition and are higher by $\sim$ 50\% for  2 -- 5. Such a behaviour by the Tayal $\Upsilon$ data appears to be anomalous in comparison with our results, particularly when there is a considerably closer agreement towards the higher end of the temperature range. Some differences between the two sets of data are expected at lower temperatures because of the position of resonances, a slight shift of which may alter the values of $\Upsilon$.  However,  large discrepancies observed for these three (and many other) transitions are not normally possible unless very high and broad resonances are present (or absent)  close to the threshold, which does not appear to be the case.  \cite{sst2} does not  show resonances for the 2 -- 3 transition, but our $\Omega$ are similar to that in  Fig. 5 for  2 -- 4 (see also Fig. 2 of \cite{sst2}), except that the magnitudes are (approximately) half for both the background as well as the peaks.  This is fully expected and  therefore the $\Upsilon$ of \cite{sst2} are clearly anomalous for the 2 -- 3/4/5 transitions, at temperatures below 10, 000 K.

\vspace*{0.2 cm}
{\LARGE Figures 7 and 8 }
\vspace*{0.2 cm}

In Fig. 8 we show one more comparison for four transitions, namely 1 -- 12 (3s$^2$3p $^2$P$^o_{1/2}$ -- 3s$^2$3d $^2$D$_{3/2}$),  1 - 14 (3s$^2$3p $^2$P$^o_{1/2}$ -- 3s3p$^2$ $^2$P$_{1/2}$), 2-- 13 (3s$^2$3p $^2$P$^o_{3/2}$ -- 3s$^2$3d $^2$D$_{5/2}$) and 2 -- 15 (3s$^2$3p $^2$P$^o_{3/2}$ -- 3s3p$^2$ $^2$P$_{3/2}$). For all these transitions the $\Upsilon$ of \cite{sst2} are higher at all temperatures. Furthermore, the discrepancy with our calculations increases with the temperature,  although his values of $\Omega$ differ more towards the lower end of the energy range (see Fig. 3). For the 1 -- 14, 2 -- 13 and 2 -- 15 transitions  the f- values calculated from GRASP and MCHF are large and  (probably) similar (see Table 3), and yet the Tayal values of $\Upsilon$ are higher by $\sim$ 50\%. Finally, we note that comparisons shown  in  Fig. 8 of  \cite{sst2} with the $\Upsilon$ of \cite{dk1} for the  3s3p$^2$ $^4$P$_{1/2,3/2}$ -- 3s$^2$3d $^2$D$_{3/2}$ (i.e. 3/4 -- 12)  forbidden transitions must be incorrect, because the latter did not report results for these.  This is likely because \cite{dk1} mistakenly labelled the 3s3p$^2$  $^2$D$_{3/2,5/2}$ levels as 3s$^2$3d $^2$D$_{3/2,5/2}$, but subsequently rectified this in a later paper \citep{dk3}. For the relevant 3s3p$^2$ $^4$P$_{1/2,3/2}$ -- 3s3p$^2$  $^2$D (3/4 -- 6) transitions there are no discrepancies between the two calculations, as already shown in Table 6. However, in comparison to our calculations,  the  $\Upsilon$ of \cite{sst2} for the 3/4 -- 12 transitions are overestimated by about a factor of two over the entire range of temperature.

There is comparatively a better agreement (within 20\%) between our results of $\Upsilon$ and those of \cite{mab}. However, their data are too limited for a thorough comparison. Similarly, the earlier results of \cite{dk1} are comparable with ours, except for a few transitions, such as: 1 -- 6/12, 2 -- 7/12 and 6 -- 7. Their higher  $\Upsilon$ values for these transitions are a direct consequence of their larger f- values. Nevertheless, the limitations of their data  have already been pointed in section 1. Finally, as already noted in section 1,  \cite{jch} adopted the data of \cite{dk1} in analysing UV lines of Si II and found discrepancies between theory and observations. Based on this, they estimated that the $\Upsilon$ of \cite{dk1} for the 3s$^2$3p $^2$P$^o$ -- 3s3p$^2$ $^4$P multiplet may be overestimated by a factor of 1.5, but this is not supported by any calculations performed to date, including the present one. However, the \cite{dk1}  $\Upsilon$ for the 3s$^2$3p $^2$P$^o$ -- 3s3p$^2$ $^2$D multiplet do appear to be overestimated by nearly the same factor.

\section{Conclusions}

Energies and lifetimes for the lowest 56 levels of  Si II  belonging to the $n \le$ 5 configurations are reported, along with radiative rates for four types of transitions (E1, E2, M1 and M2), calculated with  the {\sc grasp} code. Additionally, calculations have also been performed with the {\sc fac}  code for comparison purposes.  Based on  comparisons with measurements and available theoretical results,  our energy levels are estimated to be accurate to $\sim$ 0.1 Ryd. However, scope remains for improvement, although the inclusion of extensive CI is not very helpful in improving the accuracy further. Similarly, for a majority of (strong) transitions our listed A- values (and other related parameters including  lifetimes) are assessed to be accurate to better than 20\%.

For collision strengths $\Omega$ and effective collision strengths $\Upsilon$,  limited previous results are available for comparison. However, \cite{sst2} has reported data  over a wide range of energy and temperature,  listing values of $\Omega$ and $\Upsilon$  for 59 and 465 transitions, respectively. Our results are much more extensive considering 1540 transitions among 56 levels. Discrepancies of $\sim$ 25\% between our calculations of $\Omega$ and those of \cite{sst2} are noted for several transitions, both allowed as well as forbidden. For most transitions his results are higher, but some are lower. We have included a large range of partial waves, our results for $\Omega$ have fully converged at energies below 6 Ryd, and for higher energies the contribution of higher neglected partial waves has been taken into account. Therefore, we see no apparent deficiency in our work and estimate the accuracy of $\Omega$ to be better than 20\% for a majority of transitions. 

For  calculations of $\Upsilon$, resonances in the thresholds energy region have been resolved in a fine mesh, and are observed to be significant for many transitions.  However, differences with the corresponding results of \cite{sst2} are up to a factor of two for many transitions, and over the entire range of temperature from 10$^{3.4}$ to 10$^{5.4}$ K. For most transitions his $\Upsilon$ values are overestimated, but some are underestimated. Although Tayal included an equally large range of partial waves and energy for calculating $\Omega$ and resolved resonances to determine $\Upsilon$, the differences with our calculations are significant.  In fact the energy resolution in his calculations was finer (0.00025 Ryd) in the thresholds region, but it should not make any appreciable difference in the reported results of $\Upsilon$, because the density of resonances is not very high, as seen in Figs. 4--6. Nevertheless, to confirm it we have performed additional  calculations with a larger resolution of 0.0005 Ryd at energies below 0.525 Ryd. These calculations affect transitions among the lowest 5 levels and particularly at lower temperatures. However, differences between the two sets of calculations for all transitions are less than 0.4\% at all temperatures. Therefore,  we have confidence in our results although scope remains for improvement. We believe  that the complete set of data  presented here  for both radiative and excitation rates for transitions in  Si II  will  be useful for diagnosing and  modelling  of astrophysical plasmas, and that some of the existing discrepancies noted by \cite{jch} and \cite{bald} will be resolved. 

\section*{Acknowledgments}

KMA is thankful to  the Atomic Weapons Establishment, Aldermaston for    financial support.

\clearpage
\newpage

\setcounter{table}{0}                                                                                         
\begin{table*}                                                                                                
\caption{Energy  levels of Si II, their threshold energies (in Ryd) and lifetimes (s).  ($a{\pm}b \equiv a{\times}$10$^{{\pm}b}$). } 
\begin{tabular}{rllrrrrrrrrrr} \hline
Index & Configuration  & Level               &   NIST    & SST      &  GRASP1  &   FAC1  &  FAC2   & FAC3    & $\tau$ (s)  \\
\hline
   1  &  3s$^2$3p       &  $^2$P$^o _{1/2}$  &  0.00000  & 0.00000  & 0.00000  & 0.00000 & 0.00000 & 0.00000 & .......... \\
   2  &  3s$^2$3p       &  $^2$P$^o _{3/2}$  &  0.00262  & 0.00262  & 0.00233  & 0.00224 & 0.00194 & 0.00195 & 6.693$+$03 \\
   3  &  3s3p$^2$       &  $^4$P$   _{1/2}$  &  0.39024  & 0.38056  & 0.35314  & 0.36774 & 0.35294 & 0.35373 & 2.510$-$04 \\
   4  &  3s3p$^2$       &  $^4$P$   _{3/2}$  &  0.39123  & 0.38139  & 0.35411  & 0.36828 & 0.35370 & 0.35449 & 8.004$-$03 \\
   5  &  3s3p$^2$       &  $^4$P$   _{5/2}$  &  0.39283  & 0.38275  & 0.35569  & 0.36996 & 0.35494 & 0.35573 & 3.391$-$03 \\
   6  &  3s3p$^2$       &  $^2$D$   _{3/2}$  &  0.50402  & 0.50142  & 0.52570  & 0.53233 & 0.48520 & 0.48597 & 5.629$-$06 \\
   7  &  3s3p$^2$       &  $^2$D$   _{5/2}$  &  0.50416  & 0.50159  & 0.52583  & 0.53252 & 0.48541 & 0.48618 & 4.254$-$06 \\
   8  &  3s$^2$4s       &  $^2$S$   _{1/2}$  &  0.59688  & 0.59763  & 0.64329  & 0.66222 & 0.64309 & 0.64347 & 6.705$-$10 \\
   9  &  3s3p$^2$       &  $^2$S$   _{1/2}$  &  0.69863  & 0.70229  & 0.76675  & 0.75842 & 0.75351 & 0.75421 & 7.363$-$10 \\
  10  &  3s$^2$4p       &  $^2$P$^o _{1/2}$  &  0.73987  & 0.72724  & 0.77820  & 0.79569 & 0.78195 & 0.78164 & 8.891$-$09 \\
  11  &  3s$^2$4p       &  $^2$P$^o _{3/2}$  &  0.74042  & 0.73769  & 0.77871  & 0.79597 & 0.78211 & 0.78182 & 8.835$-$09 \\
  12  &  3s$^2$3d       &  $^2$D$   _{3/2}$  &  0.72299  & 0.72889  & 0.79914  & 0.80484 & 0.73396 & 0.73261 & 3.078$-$10 \\
  13  &  3s$^2$3d       &  $^2$D$   _{5/2}$  &  0.72314  & 0.72905  & 0.79931  & 0.80501 & 0.73406 & 0.73271 & 3.092$-$10 \\
  14  &  3s3p$^2$       &  $^2$P$   _{1/2}$  &  0.76366  & 0.77039  & 0.89104  & 0.87977 & 0.82156 & 0.82235 & 1.850$-$10 \\
  15  &  3s3p$^2$       &  $^2$P$   _{3/2}$  &  0.76550  & 0.77468  & 0.89273  & 0.88128 & 0.82280 & 0.82359 & 1.847$-$10 \\
  16  &  3s$^2$5s       &  $^2$S$   _{1/2}$  &  0.89279  & 0.89164  & 0.92186  & 0.92866 & 0.91463 & 0.91452 & 1.865$-$09 \\
  17  &  3s3p($^3$P)3d  &  $^2$D$^o _{3/2}$  &  0.99126	 & 0.98377  & 0.95559  & 0.98332 & 0.93831 & 0.93910 & 4.397$-$07 \\  
  18  &  3s3p($^3$P)3d  &  $^2$D$^o _{5/2}$  &  0.99165	 & 0.98421  & 0.95604  & 0.98356 & 0.93869 & 0.93947 & 4.319$-$07 \\  
  19  &  3s$^2$4d       &  $^2$D$   _{3/2}$  &  0.92059  & 0.92128  & 0.96139  & 0.96016 & 0.94787 & 0.94554 & 6.419$-$10 \\
  20  &  3s$^2$4d       &  $^2$D$   _{5/2}$  &  0.92060  & 0.92131  & 0.96141  & 0.96016 & 0.94784 & 0.94552 & 6.456$-$10 \\
  21  &  3s$^2$4f       &  $^2$F$^o _{7/2}$  &  0.94367  & 0.94370  & 0.96650  & 0.97091 & 0.95576 & 0.95654 & 3.881$-$09 \\
  22  &  3s$^2$4f       &  $^2$F$^o _{5/2}$  &  0.94367  & 0.94370  & 0.96650  & 0.97091 & 0.95576 & 0.95655 & 3.886$-$09 \\
  23  &  3s$^2$5p       &  $^2$P$^o _{1/2}$  &  0.94645  & 0.94287  & 0.97479  & 0.98940 & 0.97822 & 0.97662 & 1.596$-$08 \\
  24  &  3s$^2$5p       &  $^2$P$^o _{3/2}$  &  0.94667  & 0.94305  & 0.97500  & 0.98957 & 0.97824 & 0.97672 & 1.597$-$08 \\
  25  &  3s3p($^3$P)3d  &  $^4$F$^o _{3/2}$  &  1.04127  &	    & 1.00875  & 1.02983 & 0.98646 & 0.98725 & 7.061$-$06 \\  
  26  &  3s3p($^3$P)3d  &  $^4$F$^o _{5/2}$  &  1.04183  &	    & 1.00932  & 1.03366 & 0.98694 & 0.98773 & 3.746$-$06 \\  
  27  &  3s3p($^3$P)3d  &  $^4$F$^o _{7/2}$  &  1.04262  &	    & 1.01013  & 1.03593 & 0.98761 & 0.98840 & 3.601$-$06 \\  
  28  &  3s3p($^3$P)3d  &  $^4$F$^o _{9/2}$  &  1.04367  &	    & 1.01119  & 1.03355 & 0.98849 & 0.98928 & 2.129$+$01 \\  
  29  &  3s$^2$5d       &  $^2$D$   _{3/2}$  &  1.02421  & 1.02377  & 1.05396  & 1.05700 & 1.05998 & 1.04788 & 2.146$-$09 \\
  30  &  3s$^2$5d       &  $^2$D$   _{5/2}$  &  1.02422  & 1.02378  & 1.05396  & 1.05697 & 1.05987 & 1.04784 & 2.157$-$09 \\
  31  &  3s$^2$5f       &  $^2$F$^o _{7/2}$  &  1.03666  & 1.03511  & 1.05837  & 1.07196 & 1.05601 & 1.05678 & 9.149$-$09 \\
  32  &  3s$^2$5f       &  $^2$F$^o _{5/2}$  &  1.03666  & 1.03511  & 1.05837  & 1.07197 & 1.05602 & 1.05678 & 9.164$-$09 \\
  33  &  3s$^2$5g       &  $^2$G$   _{7/2}$  &  1.04046  &	    & 1.06176  & 1.06279 & 1.04711 & 1.04790 & 1.397$-$08 \\
  34  &  3s$^2$5g       &  $^2$G$   _{9/2}$  &  1.04046  &	    & 1.06176  & 1.06279 & 1.04711 & 1.04790 & 1.397$-$08 \\
  35  &  3s3p($^3$P)3d  &  $^4$P$^o _{5/2}$  &  	 &	    & 1.09742  & 1.11565 & 1.08685 & 1.08763 & 4.540$-$10 \\  
  36  &  3s3p($^3$P)3d  &  $^4$P$^o _{3/2}$  &  	 &	    & 1.09836  & 1.11493 & 1.08768 & 1.08847 & 4.520$-$10 \\  
  37  &  3s3p($^3$P)3d  &  $^4$P$^o _{1/2}$  &  	 &	    & 1.09900  & 1.11657 & 1.08824 & 1.08903 & 4.545$-$10 \\  
  38  &  3s3p($^3$P)3d  &  $^4$D$^o _{1/2}$  &  	 &	    & 1.10612  & 1.12194 & 1.09802 & 1.09881 & 2.465$-$10 \\  
  39  &  3s3p($^3$P)3d  &  $^4$D$^o _{3/2}$  &  	 &	    & 1.10640  & 1.12361 & 1.09820 & 1.09899 & 2.477$-$10 \\  
  40  &  3s3p($^3$P)3d  &  $^4$D$^o _{5/2}$  &  	 &	    & 1.10672  & 1.12461 & 1.09844 & 1.09923 & 2.479$-$10 \\  
  41  &  3s3p($^3$P)3d  &  $^4$D$^o _{7/2}$  &  	 &	    & 1.10700  & 1.12303 & 1.09866 & 1.09945 & 2.465$-$10 \\  
  42  &  3p$^3$         &  $^4$S$^o _{3/2}$  &  	 &	    & 1.10799  & 1.12719 & 1.10455 & 1.10534 & 1.973$-$10 \\
  43  &  3s3p($^3$P)3d  &  $^2$P$^o _{3/2}$  &  	 &	    & 1.15102  & 1.16203 & 1.15549 & 1.15765 & 1.404$-$09 \\  
  44  &  3s3p($^3$P)3d  &  $^2$P$^o _{1/2}$  &  	 &	    & 1.15184  & 1.16236 & 1.15602 & 1.15819 & 1.394$-$09 \\  
  45  &  3s3p($^3$P)3d  &  $^2$F$^o _{5/2}$  &  	 &	    & 1.21481  & 1.22215 & 1.22773 & 1.22857 & 4.865$-$10 \\  
  46  &  3s3p($^3$P)3d  &  $^2$F$^o _{7/2}$  &  	 &	    & 1.21707  & 1.22420 & 1.22969 & 1.23053 & 4.828$-$10 \\  
  47  &  3p$^3$         &  $^2$D$^o _{5/2}$  &  	 &	    & 1.28531  & 1.29345 & 1.25126 & 1.25204 & 2.109$-$10 \\
  48  &  3p$^3$         &  $^2$D$^o _{3/2}$  &  	 &	    & 1.28548  & 1.29386 & 1.25138 & 1.25217 & 2.109$-$10 \\
  49  &  3p$^3$         &  $^2$P$^o _{1/2}$  &  	 &	    & 1.41958  & 1.41436 & 1.42672 & 1.42754 & 2.334$-$10 \\
  50  &  3p$^3$         &  $^2$P$^o _{3/2}$  &  	 &	    & 1.41960  & 1.41475 & 1.42678 & 1.42759 & 2.330$-$10 \\
  51  &  3s3p($^1$P)3d  &  $^2$F$^o _{7/2}$  &  	 &	    & 1.47195  & 1.47076 & 1.43957 & 1.44241 & 1.843$-$10 \\  
  52  &  3s3p($^1$P)3d  &  $^2$F$^o _{5/2}$  &  	 &	    & 1.47258  & 1.47138 & 1.44027 & 1.44310 & 1.836$-$10 \\  
  53  &  3s3p($^1$P)3d  &  $^2$D$^o _{3/2}$  &  	 &	    & 1.56391  & 1.55317 & 1.54933 & 1.55012 & 1.376$-$10 \\  
  54  &  3s3p($^1$P)3d  &  $^2$D$^o _{5/2}$  &  	 &	    & 1.56428  & 1.55328 & 1.54973 & 1.55052 & 1.377$-$10 \\  
  55  &  3s3p($^1$P)3d  &  $^2$P$^o _{1/2}$  &  	 &	    & 1.60151  & 1.58768 & 1.55439 & 1.55520 & 9.231$-$11 \\  
  56  &  3s3p($^1$P)3d  &  $^2$P$^o _{3/2}$  &  	 &	    & 1.60159  & 1.58734 & 1.55438 & 1.55519 & 9.246$-$11 \\  
 \hline	
\end{tabular}

\begin{flushleft}
{\small
NIST: {\tt http://www.nist.gov/pml/data/asd.cfm} \\
SST:  \cite{sst2} \\ 
GRASP: Present results including the QED effects   \\
FAC1: Present results with 56 levels \\
FAC2: Present results with 164 levels \\
FAC3: Present results with 175 levels \\
}
\end{flushleft}
\end{table*} 


\clearpage
\newpage  

\setcounter{table}{1}                                                                                                                                           
\begin{table*}                                                                                                                                                  
\caption{Transition wavelengths ($\lambda_{ij}$ in $\AA$), radiative rates (A$_{ji}$ in s$^{-1}$), oscillator strengths (f$_{ij}$, dimensionless), and line  
strengths (S, in atomic units) for electric dipole (E1), and A$_{ji}$ for E2, M1 and M2 transitions in Si II. ($a{\pm}b \equiv a{\times}$10$^{{\pm}b}$).}     
\begin{tabular}{rrrrrrrrr}                                                                                                                                      
\hline                                                                                                                                                          
\hline                                                                                                                                                          
$i$ & $j$ & $\lambda_{ij}$ & A$^{{\rm E1}}_{ji}$  & f$^{{\rm E1}}_{ij}$ & S$^{{\rm E1}}$ & A$^{{\rm E2}}_{ji}$  & A$^{{\rm M1}}_{ji}$ & A$^{{\rm M2}}_{ji}$ \\  
\hline                                                                                                                                                   
    1 &    2 &  3.919$+$05 &  0.000$+$00 &  0.000$+$00 &  0.000$+$00 &  1.082$-$09 &  1.494$-$04 &  0.000$+$00 \\       
    1 &    3 &  2.580$+$03 &  1.703$+$03 &  1.700$-$06 &  2.889$-$05 &  0.000$+$00 &  0.000$+$00 &  0.000$+$00 \\       
    1 &    4 &  2.573$+$03 &  8.227$+$00 &  1.634$-$08 &  2.768$-$07 &  0.000$+$00 &  0.000$+$00 &  2.818$-$03 \\       
    1 &    5 &  2.562$+$03 &  0.000$+$00 &  0.000$+$00 &  0.000$+$00 &  0.000$+$00 &  0.000$+$00 &  1.103$-$03 \\       
    1 &    6 &  1.734$+$03 &  1.167$+$05 &  1.051$-$04 &  1.200$-$03 &  0.000$+$00 &  0.000$+$00 &  3.553$-$07 \\       
    1 &    7 &  1.733$+$03 &  0.000$+$00 &  0.000$+$00 &  0.000$+$00 &  0.000$+$00 &  0.000$+$00 &  2.256$-$02 \\       
    1 &    8 &  1.417$+$03 &  4.996$+$08 &  1.503$-$01 &  1.402$+$00 &  0.000$+$00 &  0.000$+$00 &  0.000$+$00 \\       
    1 &    9 &  1.188$+$03 &  4.781$+$08 &  1.012$-$01 &  7.923$-$01 &  0.000$+$00 &  0.000$+$00 &  0.000$+$00 \\       
    1 &   10 &  1.171$+$03 &  0.000$+$00 &  0.000$+$00 &  0.000$+$00 &  0.000$+$00 &  8.890$-$06 &  0.000$+$00 \\       
    .  &      .  &  .                      &   . & . &. & . & .  & .   \\
    .  &      .  &  .                      &   . & . &. & . & .  & .   \\
    .  &      .  &  .                      &   . & . &. & . & .  & .   \\
     
\hline                                                                                                                                                          
\end{tabular}                                                                                                                                                   
\end{table*}                                  

\clearpage
\newpage  

\setcounter{table}{2}                                                                                        
\begin{table*}                                                                                                
\caption{Comparison of oscillator strengths (f- values) for transitions among the lowest 30 levels of Si II.  ($a{\pm}b \equiv a{\times}$10$^{{\pm}b}$).}            
{\tiny
\begin{tabular}{rrrrrrrrrrrrrrrrrr} \hline
I     &    J & GRASP      & MCHF       &    AS      & R        &  I  &    J & GRASP      & MCHF       & AS          & R          \\
\hline
    1 &    3 &  1.700$-$6 &  ......... &  4.500$-$6 &  1.8$-$1 &   9 &   23 &  1.715$-$4 &  ......... &  .........  &	2.5$+$0  \\
    1 &    4 &  1.634$-$8 &  ......... &  2.000$-$9 &  4.5$+$0 &   9 &   24 &  3.588$-$4 &  ......... &  .........  &	2.5$+$0  \\
    1 &    6 &  1.051$-$4 &  2.500$-$3 &  2.400$-$3 &  3.2$+$1 &   9 &   25 &  7.61$-$10 &  ......... &  .........  &	1.3$+$0  \\
    1 &    8 &  1.503$-$1 &  1.310$-$1 &  1.400$-$1 &  5.6$-$1 &  10 &   12 &  9.477$-$2 &  1.680$-$2 &  .........  &	4.6$-$1  \\
    1 &    9 &  1.012$-$1 &  9.210$-$2 &  9.100$-$2 &  7.8$-$1 &  10 &   14 &  2.559$-$4 &  ......... &  .........  &	2.3$+$0  \\
    1 &   12 &  1.063$+$0 &  1.180$+$0 &  1.200$+$0 &  8.6$-$1 &  10 &   15 &  2.730$-$4 &  ......... &  .........  &	1.8$+$0  \\
    1 &   14 &  5.616$-$1 &  5.780$-$1 &  5.400$-$1 &  8.7$-$1 &  10 &   16 &  2.392$-$1 &  2.370$-$1 &  .........  &	9.3$-$1  \\
    1 &   15 &  2.789$-$1 &  2.850$-$1 &  2.600$-$1 &  8.7$-$1 &  10 &   19 &  8.121$-$1 &  9.090$-$1 &  .........  &	1.0$+$0  \\
    1 &   16 &  1.942$-$2 &  1.490$-$2 &  ......... &  5.5$-$1 &  10 &   29 &  1.665$-$1 &  1.060$-$1 &  .........  &	9.9$-$1  \\
    1 &   19 &  3.184$-$1 &  1.700$-$1 &  ......... &  6.7$-$1 &  11 &   12 &  9.264$-$3 &  3.500$-$3 &  .........  &	4.7$-$1  \\
    1 &   29 &  6.900$-$2 &  4.170$-$2 &  ......... &  1.0$+$0 &  11 &   13 &  8.404$-$2 &  2.070$-$2 &  .........  &	4.7$-$1  \\
    2 &    3 &  1.154$-$6 &  ......... &  1.700$-$6 &  2.0$-$1 &  11 &   14 &  1.764$-$5 &  ......... &  .........  &	5.2$+$0  \\
    2 &    4 &  1.174$-$7 &  ......... &  1.000$-$6 &  2.1$-$1 &  11 &   15 &  1.661$-$4 &  ......... &  .........  &	3.2$+$0  \\
    2 &    5 &  4.410$-$7 &  ......... &  2.600$-$6 &  1.2$+$0 &  11 &   16 &  2.406$-$1 &  2.380$-$1 &  .........  &	9.3$-$1  \\
    2 &    6 &  2.771$-$5 &  1.700$-$4 &  1.800$-$4 &  9.9$+$0 &  11 &   19 &  8.158$-$2 &  9.110$-$2 &  .........  &	1.0$+$0  \\
    2 &    7 &  1.602$-$4 &  1.970$-$3 &  1.700$-$3 &  1.7$+$1 &  11 &   20 &  7.333$-$1 &  8.190$-$1 &  .........  &	1.0$+$0  \\
    2 &    8 &  1.503$-$1 &  1.300$-$1 &  1.300$-$1 &  5.6$-$1 &  11 &   29 &  1.663$-$2 &  1.040$-$2 &  .........  &	9.9$-$1  \\
    2 &    9 &  9.374$-$2 &  8.090$-$2 &  8.100$-$2 &  7.8$-$1 &  11 &   30 &  1.496$-$1 &  9.340$-$2 &  .........  &	9.9$-$1  \\
    2 &   12 &  1.025$-$1 &  1.110$-$1 &  1.000$-$1 &  8.6$-$1 &  12 &   17 &  1.447$-$5 &  ......... &  .........  &	8.4$+$0  \\
    2 &   13 &  9.506$-$1 &  1.050$+$0 &  1.000$+$0 &  8.6$-$1 &  12 &   18 &  8.911$-$6 &  ......... &  .........  &	8.5$+$0  \\
    2 &   14 &  1.437$-$1 &  1.510$-$1 &  1.400$-$1 &  8.7$-$1 &  12 &   22 &  7.344$-$1 &  ......... &  .........  &	7.6$-$1  \\
    2 &   15 &  7.100$-$1 &  7.390$-$1 &  4.800$-$1 &  8.7$-$1 &  12 &   23 &  2.702$-$2 &  6.890$-$3 &  .........  &	2.2$-$1  \\
    2 &   16 &  2.097$-$2 &  1.520$-$2 &  ......... &  5.6$-$1 &  12 &   24 &  5.313$-$3 &  8.140$-$3 &  .........  &	2.2$-$1  \\
    2 &   19 &  3.310$-$2 &  1.750$-$2 &  ......... &  6.7$-$1 &  12 &   25 &  1.631$-$8 &  ......... &  .........  &	1.8$+$1  \\
    2 &   20 &  2.878$-$1 &  1.520$-$1 &  ......... &  6.7$-$1 &  12 &   26 &  1.204$-$5 &  ......... &  .........  &	8.7$-$1  \\
    2 &   29 &  7.024$-$3 &  4.180$-$3 &  ......... &  1.0$+$0 &  13 &   17 &  4.734$-$9 &  ......... &  .........  &	7.2$+$3  \\
    2 &   30 &  6.214$-$2 &  3.740$-$2 &  ......... &  1.0$+$0 &  13 &   18 &  2.262$-$5 &  ......... &   ......... &	5.3$+$0  \\
    3 &   10 &  1.382$-$7 &  ......... &  ......... &  4.8$-$1 &  13 &   21 &  6.992$-$1 &  ......... &   ......... &	7.6$-$1  \\
    3 &   11 &  1.885$-$7 &  ......... &  1.500$-$7 &  4.4$-$1 &  13 &   22 &  3.495$-$2 &  ......... &   ......... &	7.6$-$1  \\
    3 &   17 &  4.348$-$6 &  ......... &  ......... &  8.3$-$1 &  13 &   24 &  3.210$-$2 &  ......... &   ......... &	2.2$-$1  \\
    3 &   23 &  1.774$-$7 &  ......... &  ......... &  2.9$-$1 &  13 &   25 &  7.97$-$10 &  ......... &   ......... &	1.8$+$1  \\
    3 &   24 &  8.658$-$8 &  ......... &  ......... &  9.2$-$2 &  13 &   26 &  4.425$-$7 &  ......... &   ......... &	2.7$+$0  \\
    3 &   25 &  3.017$-$5 &  ......... &  ......... &  7.2$-$1 &  13 &   27 &  1.509$-$5 &  ......... &   ......... &	9.5$-$1  \\
    4 &   10 &  1.220$-$7 &  ......... &  3.000$-$7 &  2.5$-$1 &  14 &   17 &  7.666$-$5 &  ......... &   ......... &	1.9$+$1  \\
    4 &   11 &  3.396$-$9 &  ......... &  2.900$-$8 &  2.0$-$2 &  14 &   23 &  8.043$-$8 &  ......... &   ......... &	2.0$+$1  \\
    4 &   17 &  7.207$-$7 &  ......... &  ......... &  9.9$-$1 &  14 &   24 &  8.815$-$5 &  ......... &   ......... &	2.3$-$1  \\
    4 &   18 &  9.915$-$6 &  ......... &  ......... &  8.2$-$1 &  14 &   25 &  1.628$-$7 &  ......... &   ......... &	7.2$+$0  \\
    4 &   22 &  6.561$-$7 &  ......... &  ......... &  9.5$-$1 &  15 &   17 &  5.351$-$6 &  ......... &   ......... &	6.4$+$0  \\
    4 &   23 &  6.962$-$8 &  ......... &  ......... &  2.7$-$2 &  15 &   18 &  7.031$-$5 &  ......... &   ......... &	1.5$+$1  \\
    4 &   24 &  4.594$-$8 &  ......... &  ......... &  1.3$+$0 &  15 &   22 &  4.104$-$5 &  ......... &   ......... &	9.3$-$1  \\
    4 &   25 &  1.898$-$5 &  ......... &  ......... &  7.2$-$1 &  15 &   23 &  1.158$-$6 &  ......... &   ......... &	2.3$+$0  \\
    4 &   26 &  7.766$-$5 &  ......... &  ......... &  7.2$-$1 &  15 &   24 &  4.258$-$5 &  ......... &   ......... &	1.7$-$1  \\
    5 &   11 &  7.569$-$7 &  ......... &  1.800$-$6 &  2.1$-$1 &  15 &   25 &  1.861$-$8 &  ......... &   ......... &	4.0$+$0  \\
    5 &   17 &  1.576$-$9 &  ......... &  ......... &  4.2$-$3 &  15 &   26 &  5.387$-$8 &  ......... &   ......... &	7.6$+$0  \\
    5 &   18 &  1.040$-$7 &  ......... &  ......... &  1.9$-$1 &  16 &   17 &  2.202$-$5 &  ......... &   ......... &	5.6$-$1  \\
    5 &   21 &  3.735$-$6 &  ......... &  ......... &  9.6$-$1 &  16 &   23 &  5.800$-$1 &  5.420$-$1 &   ......... &	9.5$-$1  \\
    5 &   22 &  1.850$-$7 &  ......... &  ......... &  9.6$-$1 &  16 &   24 &  1.163$+$0 &  1.090$+$0 &   ......... &	9.5$-$1  \\
    5 &   24 &  1.615$-$7 &  ......... &  ......... &  1.1$-$1 &  16 &   25 &  3.848$-$9 &  ......... &   ......... &	1.3$+$0  \\
    5 &   25 &  1.158$-$6 &  ......... &  ......... &  7.2$-$1 &  17 &   19 &  2.94$-$10 &  ......... &   ......... &	9.2$+$4  \\
    5 &   26 &  2.159$-$5 &  ......... &  ......... &  7.2$-$1 &  17 &   20 &  3.977$-$6 &  ......... &   ......... &	7.5$+$0  \\
    5 &   27 &  1.044$-$4 &  ......... &  ......... &  7.2$-$1 &  17 &   29 &  1.063$-$5 &  ......... &   ......... &	3.0$-$3  \\
    6 &   10 &  4.639$-$2 &  4.820$-$2 &  3.500$-$2 &  3.2$-$1 &  17 &   30 &  2.044$-$5 &  ......... &   ......... &	5.6$-$1  \\
    6 &   11 &  9.235$-$3 &  9.630$-$3 &  7.000$-$3 &  3.2$-$1 &  18 &   19 &  1.304$-$6 &  ......... &   ......... &	1.8$+$0  \\
    6 &   17 &  1.289$-$3 &  ......... &  ......... &  4.6$-$1 &  18 &   20 &  1.262$-$7 &  ......... &   ......... &	1.5$+$2  \\
    6 &   18 &  1.522$-$4 &  ......... &  ......... &  5.1$-$1 &  18 &   29 &  9.960$-$8 &  ......... &   ......... &	1.4$+$1  \\
    6 &   22 &  1.249$-$1 &  ......... &  ......... &  9.3$-$1 &  18 &   30 &  2.824$-$6 &  ......... &   ......... &	4.6$-$1  \\
    6 &   23 &  1.084$-$2 &  1.030$-$2 &  ......... &  2.9$-$2 &  19 &   22 &  5.851$-$2 &  ......... &   ......... &	7.4$+$0  \\
    6 &   24 &  2.145$-$3 &  2.060$-$3 &  ......... &  2.9$-$2 &  19 &   23 &  1.224$-$1 &  1.920$-$1 &   ......... &	1.7$+$0  \\
    6 &   25 &  8.624$-$6 &  ......... &  ......... &  5.2$-$1 &  19 &   24 &  2.482$-$2 &  3.870$-$2 &   ......... &	1.7$+$0  \\
    6 &   26 &  8.233$-$7 &  ......... &  ......... &  2.5$+$0 &  19 &   25 &  7.247$-$9 &  ......... &   ......... &	1.4$+$2  \\
    7 &   11 &  5.554$-$2 &  5.790$-$2 &  4.000$-$2 &  3.2$-$1 &  19 &   26 &  4.360$-$6 &  ......... &   ......... &	4.8$-$4  \\
    7 &   17 &  1.489$-$4 &  ......... &  ......... &  5.2$-$1 &  20 &   21 &  5.543$-$2 &  ......... &   ......... &	7.4$+$0  \\
    7 &   18 &  1.426$-$3 &  ......... &  ......... &  4.8$-$1 &  20 &   22 &  2.772$-$3 &  ......... &   ......... &	7.4$+$0  \\
    7 &   21 &  1.196$-$1 &  ......... &  ......... &  9.3$-$1 &  20 &   24 &  1.487$-$1 &  2.320$-$1 &   ......... &	1.7$+$0  \\
    7 &   22 &  6.010$-$3 &  ......... &  ......... &  9.3$-$1 &  20 &   25 &  6.19$-$11 &  ......... &   ......... &	7.4$+$2  \\
    7 &   24 &  1.297$-$2 &  1.240$-$2 &  ......... &  3.0$-$2 &  20 &   26 &  1.608$-$7 &  ......... &   ......... &	3.2$+$0  \\
    7 &   25 &  7.271$-$7 &  ......... &  ......... &  5.2$-$1 &  20 &   27 &  5.636$-$6 &  ......... &   ......... &	5.4$-$3  \\
    7 &   26 &  5.471$-$6 &  ......... &  ......... &  6.3$-$1 &  21 &   30 &  1.451$-$2 &  ......... &   ......... &	3.1$+$0  \\
    7 &   27 &  3.039$-$6 &  ......... &  ......... &  1.5$+$0 &  22 &   29 &  1.356$-$2 &  ......... &   ......... &	3.1$+$0  \\
    8 &   10 &  4.442$-$1 &  3.770$-$1 &  2.400$-$1 &  9.2$-$1 &  22 &   30 &  9.670$-$4 &  ......... &   ......... &	3.1$+$0  \\
    8 &   11 &  8.912$-$1 &  7.570$-$1 &  4.900$-$1 &  9.2$-$1 &  23 &   29 &  1.113$+$0 &  1.140$+$0 &   ......... &	1.0$+$0  \\
    8 &   17 &  7.067$-$7 &  ......... &  ......... &  2.6$+$0 &  24 &   29 &  1.118$-$1 &  1.140$-$1 &   ......... &	1.0$+$0  \\
    8 &   23 &  7.247$-$4 &  ......... &  ......... &  3.3$-$1 &  24 &   30 &  1.006$+$0 &  1.030$+$0 &   ......... &	1.0$+$0  \\
    8 &   24 &  1.625$-$3 &  ......... &  ......... &  3.6$-$1 &  25 &   29 &  1.363$-$9 &  ......... &   ......... &	3.3$+$2  \\
    8 &   25 &  1.69$-$10 &  ......... &  ......... &  1.7$+$0 &  25 &   30 &  3.612$-$9 &  ......... &   ......... &	3.5$+$0  \\
    9 &   10 &  9.830$-$5 &  1.140$-$3 &  ......... &  3.0$+$0 &  26 &   29 &  3.220$-$6 &  ......... &   ......... &	1.1$-$1  \\
    9 &   11 &  1.999$-$4 &  2.290$-$3 &  ......... &  3.1$+$0 &  26 &   30 &  2.671$-$7 &  ......... &   ......... &	1.5$+$0  \\
    9 &   17 &  1.322$-$7 &  ......... &  ......... &  5.9$+$0 &  27 &   30 &  4.814$-$6 &  ......... &   ......... &	1.7$-$1  \\			
\hline                                                                                                        
\end{tabular}   
} 
\begin {flushleft}														   
\begin{tabbing}
aaaaaaaaaaaaaaaaaaaaaaaaaaaaaaaaaaaa\= \kill
GRASP: Present results with the {\sc grasp} code \\
MCHF: \cite{sst1} \\ 
AS: \cite{mab} \\ 
R: Ratio of velocity/length form of f- values from the GRASP calculations \\
\end{tabbing}
\end {flushleft}                                                                                             
\end{table*}                                                                                                                                                                      
                                                                                                   

\clearpage
\newpage 
\setcounter{table}{3}     
\begin{table*}      
\caption{Collision strengths for transitions from the ground state 3s$^2$3p $^2$P$^o_{1/2,3/2}$ levels of   Si II. ($a{\pm}b \equiv$ $a\times$10$^{{\pm}b}$).}          
\begin{tabular}{rrlllrrlll}                                                                                   
\hline                                                                                                        
\hline                                                                                                         
\multicolumn{2}{c}{Transition} & \multicolumn{3}{c}{Energy (Ryd)} & \multicolumn{2}{c}{Transition} & \multicolumn{3}{c}{Energy (Ryd)}  \\ 					     
\hline                                                                                                        
  $i$ & $j$ &    2 & 6 & 10 &      $i$ & $j$ &    2 & 6 & 10      \\                                 
\hline                                                                                                        
  1 &  2 &  1.516$-$0 &  1.631$-$0 &  1.706$-$0 & .. & .. &  ......... &  ........  &  .........  \\
  1 &  3 &  1.696$-$1 &  3.269$-$2 &  1.398$-$2 &  2 &  3 &  1.152$-$1 &  2.129$-$2 &  9.081$-$3  \\
  1 &  4 &  2.491$-$1 &  4.744$-$2 &  2.011$-$2 &  2 &  4 &  3.200$-$1 &  5.992$-$2 &  2.535$-$2  \\
  1 &  5 &  1.539$-$1 &  2.793$-$2 &  1.172$-$2 &  2 &  5 &  6.990$-$1 &  1.330$-$1 &  5.643$-$2  \\
  1 &  6 &  4.056$-$1 &  2.391$-$1 &  2.476$-$1 &  2 &  6 &  8.683$-$1 &  4.275$-$1 &  3.815$-$1  \\
  1 &  7 &  6.541$-$1 &  3.174$-$1 &  2.783$-$1 &  2 &  7 &  1.255$-$0 &  6.791$-$1 &  6.609$-$1  \\
  1 &  8 &  1.273$-$0 &  3.102$-$0 &  4.080$-$0 &  2 &  8 &  2.557$-$0 &  6.228$-$0 &  8.192$-$0  \\
  1 &  9 &  1.205$-$0 &  2.582$-$0 &  3.316$-$0 &  2 &  9 &  2.284$-$0 &  4.868$-$0 &  6.249$-$0  \\
  1 & 10 &  8.154$-$1 &  1.297$-$0 &  1.412$-$0 &  2 & 10 &  3.461$-$1 &  3.407$-$1 &  3.594$-$1  \\
  1 & 11 &  3.439$-$1 &  3.362$-$1 &  3.542$-$1 &  2 & 11 &  1.978$-$0 &  2.936$-$0 &  3.185$-$0  \\
  1 & 12 &  6.573$-$0 &  1.761$+$1 &  2.369$+$1 &  2 & 12 &  1.670$-$0 &  3.704$-$0 &  4.883$-$0  \\
  1 & 13 &  3.307$-$1 &  2.443$-$1 &  2.451$-$1 &  2 & 13 &  1.207$+$1 &  3.182$+$1 &  4.274$+$1  \\
  1 & 14 &  3.764$-$0 &  9.824$-$0 &  1.307$+$1 &  2 & 14 &  2.065$-$0 &  5.104$-$0 &  6.743$-$0  \\
  1 & 15 &  2.005$-$0 &  4.951$-$0 &  6.539$-$0 &  2 & 15 &  9.633$-$0 &  2.489$+$1 &  3.306$+$1  \\
  1 & 16 &  1.457$-$1 &  2.706$-$1 &  3.436$-$1 &  2 & 16 &  3.089$-$1 &  5.883$-$1 &  7.492$-$1  \\
  1 & 17 &  5.261$-$1 &  9.792$-$1 &  1.081$-$0 &  2 & 17 &  5.372$-$1 &  9.583$-$1 &  1.054$-$0  \\
  1 & 18 &  3.714$-$1 &  6.584$-$1 &  7.245$-$1 &  2 & 18 &  1.224$-$0 &  2.248$-$0 &  2.477$-$0  \\
  1 & 19 &  2.155$-$0 &  5.268$-$0 &  6.799$-$0 &  2 & 19 &  6.568$-$1 &  1.204$-$0 &  1.516$-$0  \\
  1 & 20 &  1.759$-$1 &  9.273$-$2 &  8.692$-$2 &  2 & 20 &  4.052$-$0 &  9.640$-$0 &  1.241$+$1  \\
  1 & 21 &  5.271$-$1 &  6.112$-$1 &  6.402$-$1 &  2 & 21 &  2.485$-$1 &  2.370$-$1 &  2.449$-$1  \\
  1 & 22 &  7.310$-$2 &  4.255$-$2 &  4.174$-$2 &  2 & 22 &  9.650$-$1 &  1.093$-$0 &  1.143$-$0  \\
  1 & 23 &  2.382$-$1 &  3.223$-$1 &  3.465$-$1 &  2 & 23 &  1.253$-$1 &  8.869$-$2 &  9.190$-$2  \\
  1 & 24 &  1.243$-$1 &  8.662$-$2 &  8.929$-$2 &  2 & 24 &  6.017$-$1 &  7.337$-$1 &  7.854$-$1  \\
  1 & 25 &  1.203$-$1 &  1.588$-$2 &  6.042$-$3 &  2 & 25 &  3.850$-$2 &  4.815$-$3 &  1.833$-$3  \\
  1 & 26 &  1.391$-$1 &  1.828$-$2 &  6.923$-$3 &  2 & 26 &  9.898$-$2 &  1.268$-$2 &  4.790$-$3  \\
  1 & 27 &  1.073$-$1 &  1.393$-$2 &  5.245$-$3 &  2 & 27 &  2.100$-$1 &  2.724$-$2 &  1.026$-$2  \\
  1 & 28 &  6.196$-$3 &  3.973$-$4 &  1.111$-$4 &  2 & 28 &  3.904$-$1 &  5.103$-$2 &  1.925$-$2  \\
  1 & 29 &  9.823$-$1 &  1.951$-$0 &  2.383$-$0 &  2 & 29 &  3.478$-$1 &  4.573$-$1 &  5.369$-$1  \\
  1 & 30 &  1.243$-$1 &  5.176$-$2 &  4.463$-$2 &  2 & 30 &  1.874$-$0 &  3.566$-$0 &  4.342$-$0  \\
  1 & 31 &  5.319$-$2 &  2.065$-$2 &  1.980$-$2 &  2 & 31 &  6.904$-$1 &  6.004$-$1 &  6.084$-$1  \\
  1 & 32 &  3.785$-$1 &  3.400$-$1 &  3.453$-$1 &  2 & 32 &  1.776$-$1 &  1.247$-$1 &  1.252$-$1  \\
  1 & 33 &  2.593$-$2 &  1.179$-$2 &  1.032$-$2 &  2 & 33 &  2.250$-$2 &  7.269$-$3 &  6.029$-$3  \\
  1 & 34 &  1.026$-$2 &  2.449$-$3 &  1.896$-$3 &  2 & 34 &  5.032$-$2 &  2.150$-$2 &  1.866$-$2  \\
  1 & 35 &  1.514$-$1 &  1.607$-$2 &  5.748$-$3 &  2 & 35 &  1.230$-$1 &  1.337$-$2 &  4.742$-$3  \\
  1 & 36 &  5.758$-$2 &  6.034$-$3 &  2.138$-$3 &  2 & 36 &  1.254$-$1 &  1.360$-$2 &  4.862$-$3  \\
  1 & 37 &  1.173$-$2 &  1.220$-$3 &  4.165$-$4 &  2 & 37 &  7.953$-$2 &  8.582$-$3 &  3.075$-$3  \\
  1 & 38 &  5.914$-$2 &  5.047$-$3 &  1.728$-$3 &  2 & 38 &  6.633$-$2 &  5.548$-$3 &  1.899$-$3  \\
  1 & 39 &  9.609$-$2 &  8.052$-$3 &  2.742$-$3 &  2 & 39 &  1.543$-$1 &  1.310$-$2 &  4.475$-$3  \\
  1 & 40 &  1.094$-$1 &  8.938$-$3 &  3.010$-$3 &  2 & 40 &  2.663$-$1 &  2.274$-$2 &  7.764$-$3  \\
  1 & 41 &  1.155$-$1 &  9.352$-$3 &  3.155$-$3 &  2 & 41 &  3.872$-$1 &  3.295$-$2 &  1.123$-$2  \\
  1 & 42 &  5.327$-$3 &  9.484$-$4 &  4.667$-$4 &  2 & 42 &  1.077$-$2 &  1.753$-$3 &  7.740$-$4  \\
  1 & 43 &  3.082$-$1 &  6.842$-$1 &  7.961$-$1 &  2 & 43 &  4.635$-$1 &  9.708$-$1 &  1.110$-$0  \\
  1 & 44 &  7.311$-$2 &  1.339$-$1 &  1.462$-$1 &  2 & 44 &  3.127$-$1 &  6.934$-$1 &  8.066$-$1  \\
  1 & 45 &  5.074$-$1 &  1.401$-$0 &  1.688$-$0 &  2 & 45 &  3.303$-$1 &  4.249$-$1 &  5.009$-$1  \\
  1 & 46 &  1.420$-$1 &  1.593$-$2 &  1.077$-$2 &  2 & 46 &  9.718$-$1 &  2.416$-$0 &  2.905$-$0  \\
  1 & 47 &  9.112$-$2 &  4.021$-$2 &  2.772$-$2 &  2 & 47 &  1.843$-$1 &  1.104$-$1 &  8.074$-$2  \\
  1 & 48 &  6.162$-$2 &  4.431$-$2 &  3.334$-$2 &  2 & 48 &  1.225$-$1 &  5.683$-$2 &  3.953$-$2  \\
  1 & 49 &  3.357$-$2 &  3.645$-$2 &  4.160$-$2 &  2 & 49 &  2.274$-$1 &  2.521$-$1 &  1.961$-$1  \\
  1 & 50 &  2.314$-$1 &  2.571$-$1 &  2.003$-$1 &  2 & 50 &  2.910$-$1 &  3.204$-$1 &  2.754$-$1  \\													  
  1 & 51 &  3.930$-$1 &  4.732$-$2 &  2.775$-$2 &  2 & 51 &  2.018$-$0 &  1.922$-$0 &  1.579$-$0  \\
  1 & 52 &  1.008$-$0 &  1.097$-$0 &  9.057$-$1 &  2 & 52 &  7.970$-$1 &  3.752$-$1 &  2.953$-$1  \\
  1 & 53 &  5.496$-$1 &  7.028$-$1 &  5.942$-$1 &  2 & 53 &  5.946$-$1 &  7.088$-$1 &  5.929$-$1  \\
  1 & 54 &  4.033$-$1 &  4.761$-$1 &  3.982$-$1 &  2 & 54 &  1.313$-$0 &  1.644$-$0 &  1.385$-$0  \\
  1 & 55 &  9.842$-$2 &  2.839$-$1 &  3.168$-$1 &  2 & 55 &  9.009$-$2 &  1.494$-$1 &  1.637$-$1  \\
  1 & 56 &  8.456$-$2 &  1.378$-$1 &  1.520$-$1 &  2 & 56 &  2.935$-$1 &  7.296$-$1 &  8.093$-$1  \\
\hline                                                                                                        
\end{tabular}                                                                                                 
\end{table*}    
                                                                                               
\newpage                                                                                                      
\clearpage 

\setcounter{table}{4}                                                                                                                       
\begin{table*}                                                                                                                              
\caption{Effective collision strengths for transitions in  Si II. ($a{\pm}b \equiv a{\times}10^{{\pm}b}$).}                                
\begin{tabular}{rrlllllllllr}                                                                                                               
\hline                                                                                                                                      
\hline                                                                                                                                      
\multicolumn {2}{c}{Transition} & \multicolumn{10}{c}{Temperature (log T$_e$, K)}\\                                                         
\hline                                                                                                                                      
$i$ & $j$ &  3.70 &      3.90 &      4.10 &      4.30 &     4.50 &      4.70 &      4.900   &  5.10 &      5.30 &      5.50 \\                             
\hline                                                                                                                                                        
  1 &  2 &  5.230$+$0 &  5.311$+$0 &  5.354$+$0 &  5.328$+$0 &  5.168$+$0 &  4.809$+$0 &  4.267$+$0 &  3.647$+$0 &  3.070$+$0 &  2.605$+$0 \\
  1 &  3 &  4.582$-$1 &  4.699$-$1 &  4.706$-$1 &  4.667$-$1 &  4.568$-$1 &  4.334$-$1 &  3.937$-$1 &  3.419$-$1 &  2.842$-$1 &  2.264$-$1 \\
  1 &  4 &  7.449$-$1 &  7.458$-$1 &  7.372$-$1 &  7.276$-$1 &  7.124$-$1 &  6.760$-$1 &  6.126$-$1 &  5.293$-$1 &  4.374$-$1 &  3.463$-$1 \\
  1 &  5 &  5.558$-$1 &  5.535$-$1 &  5.476$-$1 &  5.495$-$1 &  5.522$-$1 &  5.337$-$1 &  4.845$-$1 &  4.135$-$1 &  3.347$-$1 &  2.591$-$1 \\
  1 &  6 &  1.879$+$0 &  1.911$+$0 &  1.888$+$0 &  1.794$+$0 &  1.630$+$0 &  1.417$+$0 &  1.187$+$0 &  9.689$-$1 &  7.784$-$1 &  6.229$-$1 \\
  1 &  7 &  2.161$+$0 &  2.262$+$0 &  2.305$+$0 &  2.252$+$0 &  2.096$+$0 &  1.865$+$0 &  1.601$+$0 &  1.340$+$0 &  1.101$+$0 &  8.937$-$1 \\
  1 &  8 &  1.121$+$0 &  1.215$+$0 &  1.223$+$0 &  1.157$+$0 &  1.060$+$0 &  9.837$-$1 &  9.747$-$1 &  1.063$+$0 &  1.264$+$0 &  1.574$+$0 \\
  1 &  9 &  8.780$-$1 &  8.984$-$1 &  9.064$-$1 &  9.053$-$1 &  9.048$-$1 &  9.185$-$1 &  9.654$-$1 &  1.065$+$0 &  1.233$+$0 &  1.472$+$0 \\
  1 & 10 &  6.643$-$1 &  6.341$-$1 &  6.014$-$1 &  5.682$-$1 &  5.437$-$1 &  5.414$-$1 &  5.733$-$1 &  6.424$-$1 &  7.414$-$1 &  8.556$-$1 \\
    .  &      .  &  .                      &   . & . &. & . & .  & . &   . & . &.   \\
    .  &      .  &  .                      &   . & . &. & . & .  & .  &   . & . &.   \\
    .  &      .  &  .                      &   . & . &. & . & .  & .  &   . & . &.   \\ 
 
\hline                                                                                                                                      
\end{tabular}                                                                                                                               
\end{table*}                                                         

\newpage                                                                                                      
\clearpage 
   
\setcounter{table}{5}                                                                                         
\begin{table*}                                                                                                
\caption{Comparison of effective collision strengths ($\Upsilon$) for transitions from the ground  level 3s$^2$3p $^2$P$^o_{1/2}$ of Si II at three temperatures of 5000, 10 000, 20 000 K.  ($a{\pm}b \equiv a{\times}$10$^{{\pm}b}$).}
\begin{tabular}{rrrllllllllllll} \hline
I     &    J & \multicolumn{3}{c}{DARC} &    \multicolumn{3}{c}{BPRM} &    \multicolumn{3}{c}{BSRM} &   \multicolumn{3}{c}{RM} &      \\
\cline{3-14}
\multicolumn{2}{c}{T$_e$, $^{\rm o}$K} & 5K & 10K & 20K & 5K & 10K & 20K & 5K & 10K & 20K & 5K & 10K & 20K \\
\hline
    1  &    2  &  5.230  &  5.338  &  5.328  &  4.55   & 4.45   & 4.42   & 6.19   & 6.09  &  5.97   & 5.60   & 5.70   & 5.77   \\
    1  &    3  &  0.458  &  0.471  &  0.467  &  0.401  & 0.398  & 0.392  & 0.512  & 0.515 &  0.502  & 0.550  & 0.516  & 0.466  \\
    1  &    4  &  0.745  &  0.742  &  0.728  &  0.612  & 0.609  & 0.602  & 0.812  & 0.789 &  0.769  & 0.832  & 0.780  & 0.706  \\
    1  &    5  &  0.556  &  0.550  &  0.550  &  0.441  & 0.458  & 0.477  & 0.615  & 0.595 &  0.589  & 0.571  & 0.534  & 0.488  \\
    1  &    6  &  1.879  &  1.907  &  1.793  &  1.82   & 1.82   & 1.75   & 2.77   & 2.74  &  2.50   & 2.76   & 2.74   & 2.58   \\
    1  &    7  &  2.161  &  2.294  &  2.252  &  2.05   & 2.14   & 2.14   & 2.94   & 2.98  &  2.80   & 2.45   & 2.44   & 2.30   \\
    1  &    8  &  1.120  &  1.230  &  1.156  &  0.910  & 0.865  & 0.857  & 1.02   & 1.06  &  0.979  & 1.24   & 1.20   & 1.04   \\
    1  &    9  &  0.878  &  0.904  &  0.905  &  0.887  & 0.899  & 0.916  & 1.02   & 0.988 &  0.988  & 0.716  & 0.840  & 0.902  \\			
    1  &   10  &  0.665  &  0.618  &  0.568  &  .....  & .....  & .....  &  0.540 & 0.535 &  0.517  & 0.591  & 0.612  & 0.640  \\
    1  &   11  &  0.819  &  0.737  &  0.645  &  .....  & .....  & .....  &  0.561 & 0.546 &  0.509  & 0.695  & 0.682  & 0.654  \\
    1  &   12  &  2.214  &  2.364  &  2.556  &  .....  & .....  & .....  &  3.43  & 3.46  &  3.73   & 3.16   & 3.38   & 3.77   \\
    1  &   13  &  0.796  &  0.798  &  0.736  &  .....  & .....  & .....  &  1.11  & 0.993 &  0.891  & 1.19   & 1.09   & 0.981  \\
    1  &   14  &  1.792  &  1.911  &  2.099  &  .....  & .....  & .....  &  2.51  & 2.60  &  2.82   & 1.85   & 1.93   & 2.09   \\
    1  &   15  &  1.147  &  1.209  &  1.288  &  .....  & .....  & .....  &  1.64  & 1.68  &  1.76   & 1.32   & 1.35   & 1.40   \\
    2  &    3  &  0.405  &  0.401  &  0.400  &  .....  & .....  & .....  & 0.100  & 0.345 &  0.384  & 0.433  & 0.402  & 0.365  \\
    2  &    4  &  1.049  &  1.038  &  1.025  &  .....  & .....  & .....  & 0.899  & 1.07  &  1.08   & 1.13   & 1.05   & 0.956  \\
    2  &    5  &  2.046  &  2.071  &  2.048  &  .....  & .....  & .....  & 2.57   & 2.32  &  2.23   & 2.32   & 2.19   & 1.99   \\
    2  &    6  &  2.948  &  3.119  &  3.052  &  .....  & .....  & .....  & 4.03   & 4.15  &  3.88   & 3.50   & 3.48   & 3.32   \\
    2  &    7  &  5.116  &  5.264  &  5.020  &  .....  & .....  & .....  & 7.39   & 7.38  &  6.77   & 6.88   & 6.79   & 6.30   \\
    2  &    8  &  2.239  &  2.467  &  2.325  &  .....  & .....  & .....  & 1.92   & 2.14  &  1.97   & 2.49   & 2.41   & 2.15   \\
    2  &    9  &  1.753  &  1.799  &  1.784  &  .....  & .....  & .....  & 2.10   & 1.93  &  1.87   & 1.43   & 1.69   & 1.83   \\
    2  &   10  &  0.818  &  0.740  &  0.650  &  .....  & .....  & .....  &  0.545 & 0.540 &  0.507  & 0.699  & 0.687  & 0.658  \\
    2  &   11  &  2.317  &  2.096  &  1.927  &  .....  & .....  & .....  &  1.66  & 1.63  &  1.55   & 1.87   & 1.90   & 1.94   \\
    2  &   12  &  1.352  &  1.398  &  1.371  &  .....  & .....  & .....  &  1.92  & 1.80  &  1.74   & 2.13   & 2.12   & 2.21   \\
    2  &   13  &  4.588  &  4.899  &  4.252  &  .....  & .....  & .....  &  6.77  & 6.82  &  7.27   & 6.47   & 6.75   & 7.30   \\
    2  &   14  &  1.220  &  1.258  &  1.329  &  .....  & .....  & .....  &  1.91  & 1.85  &  1.89   & 1.31   & 1.34   & 1.41   \\
    2  &   15  &  4.730  &  5.066  &  5.538  &  .....  & .....  & .....  &  6.55  & 6.83  &  7.44   & 4.79   & 4.90   & 5.16   \\
    3  &    4  &  4.371  &  3.965  &  3.612  &  .....  & .....  & .....  &  3.09  & 3.28  &  3.24   & 4.92   & 4.51   & 3.94   \\
    3  &    5  &  2.300  &  2.432  &  2.474  &  .....  & .....  & .....  &  2.69  & 2.61  &  2.41   & 1.68   & 1.67   & 1.57   \\
    3  &    6  &  1.091  &  1.019  &  0.937  &  .....  & .....  & .....  &  1.26  & 1.20  &  1.08   & 1.20   & 1.20   & 1.09   \\
    3  &    7  &  0.596  &  0.584  &  0.569  &  .....  & .....  & .....  &  0.746 & 0.726 &  0.662  & 0.648  & 0.653  & 0.613  \\
    4  &    5  &  7.446  &  7.284  &  7.041  &  .....  & .....  & .....  &  7.16  & 7.07  &  6.72   & 7.36   & 6.94   & 6.31   \\
    4  &    6  &  1.677  &  1.581  &  1.469  &  .....  & .....  & .....  &  1.84  & 1.77  &  1.63   & 1.86   & 1.85   & 1.69   \\
    4  &    7  &  1.708  &  1.632  &  1.548  &  .....  & .....  & .....  &  1.70  & 1.66  &  1.57   & 1.86   & 1.86   & 1.72   \\
    5  &    6  &  1.286  &  1.252  &  1.211  &  .....  & .....  & .....  &  1.49  & 1.49  &  1.39   & 1.39   & 1.40   & 1.31   \\
    5  &    7  &  3.813  &  3.580  &  3.314  &  .....  & .....  & .....  &  4.07  & 3.86  &  3.53   & 4.17   & 4.16   & 3.80   \\
    6  &    7  &  4.364  &  4.502  &  4.701  &  .....  & .....  & .....  &  6.94  & 6.55  &  5.83   & 6.04   & 5.92   & 5.75   \\
 \hline
\end{tabular} 
\begin{flushleft}
{\small
DARC: Present results with the {\sc darc} code \\
BPRM: \cite{mab} \\ 
BSRM: \cite{sst2} \\ 
RM: \cite{dk1} \\ 
}
\end{flushleft}
\end{table*}


\newpage                                                                                                      
\clearpage 

\begin{center}
{\Large\bf \underline {Captions for Figures}}
\end{center}
\vspace*{0.5 cm}
\begin{flushleft}

Fig. 1. Partial collision strengths at three energies of  2 Ryd (circles), 6 Ryd (triangles) and 12 Ryd (stars) for three transitions of Si II, namely (a):  1 -- 12 (3s$^2$3p $^2$P$^o_{1/2}$ -- 3s$^2$3d $^2$D$_{3/2}$), (b)  2 -- 13 (3s$^2$3p $^2$P$^o_{3/2}$  -- 3s$^2$3d $^2$D$_{5/2}$) and (c):  2 -- 15 (3s$^2$3p $^2$P$^o_{3/2}$ -- 3s3p$^2$ $^2$P$_{3/2}$). \\

Fig. 2. Comparison of collision strengths  from our calculations from {\sc darc} (continuous curves) and of \cite{sst2} (broken curves) for the 1 -- 10 (circles: 3s$^2$3p $^2$P$^o_{1/2}$ -- 3s$^2$4p $^2$P$^o_{1/2}$), 1 -- 21 (triangles: 3s$^2$3p $^2$P$^o_{1/2}$ -- 3s$^2$4f $^2$F$^o_{7/2}$) and 2 -- 11 (stars: 3s$^2$3p $^2$P$^o_{3/2}$ -- 3s$^2$4p $^2$P$^o_{3/2}$) {\em forbidden} transitions of Si II. 

Fig. 3. Comparison of collision strengths  from our calculations from {\sc darc} (continuous curves) and  of \cite{sst2} (broken curves) for the 1 -- 12 (circles: 3s$^2$3p $^2$P$^o_{1/2}$ -- 3s$^2$3d $^2$D$_{3/2}$), 2 --13 (triangles: 3s$^2$3p $^2$P$^o_{3/2}$ -- 3s$^2$3d $^2$D$_{5/2}$) and 2 -- 15 (stars: 3s$^2$3p $^2$P$^o_{3/2}$ -- 3s3p$^2$ $^2$P$_{3/2}$) {\em allowed} transitions of Si II. 

Fig. 4. Collision strengths for the  1 -- 2 (3s$^2$3p $^2$P$^o_{1/2}$ -- 3s$^2$3p $^2$P$^o_{3/2}$)  transition of Si II.

Fig. 5. Collision strengths for the 2 -- 4 (3s$^2$3p $^2$P$^o_{3/2}$ -- 3s3p$^2$ $^4$P$_{3/2}$)    transition of Si II.

Fig. 6. Collision strengths for the  2 -- 7 (3s$^2$3p $^2$P$^o_{3/2}$ -- 3s3p$^2$ $^2$D$_{5/2}$)  transition of Si II.

Fig. 7.  Comparison of effective collision strengths  from our calculations from {\sc darc} (continuous curves) and of \cite{sst2} (broken curves) for the 2 -- 3 (circles: 3s$^2$3p $^2$P$^o_{3/2}$ -- 3s3p$^2$ $^4$P$_{1/2}$), 2 -- 4 (triangles: 3s$^2$3p $^2$P$^o_{3/2}$ -- 3s3p$^2$ $^4$P$_{3/2}$) and 2 --5 (stars: 3s$^2$3p $^2$P$^o_{3/2}$ -- 3s3p$^2$ $^4$P$_{5/2}$) transitions of Si II. 

Fig. 8. Comparison of effective collision strengths  from our calculations from {\sc darc} (continuous curves) and of \cite{sst2} (broken curves) for the 1 -- 12 (circles: 3s$^2$3p $^2$P$^o_{1/2}$ -- 3s$^2$3d $^2$D$_{3/2}$), 1 -- 14 (triangles: 3s$^2$3p $^2$P$^o_{1/2}$ -- 3s3p$^2$ $^2$P$_{1/2}$),  2 --13 (stars: 3s$^2$3p $^2$P$^o_{3/2}$ -- 3s$^2$3d $^2$D$_{5/2}$)  and 2 -- 15 (diamonds: 3s$^2$3p $^2$P$^o_{3/2}$ -- 3s3p$^2$ $^2$P$_{3/2}$) transitions of Si II.

\end{flushleft}

\end{document}